\documentclass[prx,amsmath,amssymb, notitlepage, twocolumn,
nofootinbib,
superscriptaddress,
longbibliography
]{revtex4-1}

\usepackage{amsmath}
\usepackage{mathrsfs}
\usepackage{tabularx,graphicx}
\usepackage{epstopdf}
\usepackage{makecell}
\usepackage{graphicx}
\usepackage{latexsym}
\usepackage{amssymb}
\usepackage{amsfonts}
\usepackage{amsmath}
\usepackage{color, colortbl}
\usepackage{psfrag}
\usepackage{bbm}
\usepackage{bm}
\usepackage{titlesec}
\usepackage{dsfont}
\usepackage{feynmp}
\usepackage{slashed}
\usepackage{multirow}
\usepackage{url}
\usepackage{wasysym}

\newcommand{\mc}[1]{\mathcal{#1}}

\newcommand{\beq}{\begin{eqnarray}}
	\newcommand{\eeq}{\end{eqnarray}}

\newcommand{\la}{\langle}
\newcommand{\ra}{\rangle}

\newcommand{\bsp}{\begin{aligned}}
	\newcommand{\esp}{\end{aligned}}

\newcommand{\ie}{{i.e., }}
\newcommand{\eg}{{e.g., }}

\usepackage{xcolor}
\definecolor{darkblue}{rgb}{0.,0.,0.4}
\definecolor{darkred}{rgb}{0.5,0.,0.}
\definecolor{BlueViolet}{RGB}{138,43,226}
\definecolor{SkyBlue}{RGB}{30,144,255}
\definecolor{DarkGreen}{RGB}{0,100,0}

\newcommand{\bk}{{\bf k}}

\newcommand{\ba}{{\bf a}}

\newcommand{\bb}{{\bf b}}

\newcommand{\br}{{\bf r}}

\newcommand{\bT}{{\bf T}}

\usepackage[colorlinks=true,linkcolor=darkblue,citecolor=blue,urlcolor=darkred]{hyperref}
\usepackage[normalem]{ulem}

\usepackage{mathrsfs}
\usepackage{comment}

\newcommand{\z}{\mathbb{Z}}
\newcommand{\R}{\mathbb{R}}
\def\U{\mathrm{U}(1)}
\usepackage{pst-all}
\usepackage{leftidx}
\usepackage[off]{auto-pst-pdf} 
\usepackage{booktabs}
\usepackage{yfonts}
\makeatletter

\usepackage[caption=false]{subfig}

\usepackage{enumitem}

\begin{document}

\hfill MIT-CTP/5720
\title{
When does a lattice higher-form symmetry flow to\\
a topological higher-form symmetry at low energies?
}
 
\author{Ruizhi Liu}
\affiliation{Perimeter Institute for Theoretical Physics, Waterloo, Ontario, Canada N2L 2Y5}
\affiliation{Department of Mathematics and Statistics, Dalhousie University, Halifax, Nova Scotia, Canada, B3H 4R2}

\author{Pok Man Tam}
\affiliation{Princeton Center for Theoretical Science, Princeton University, Princeton, NJ 08544}

\author{Ho Tat Lam}
\affiliation{Department of Physics, Massachusetts Institute of Technology, Cambridge, Massachusetts 02139, USA}
\affiliation{Leinweber Institute for Theoretical Physics,
Stanford University, Stanford, CA 94305, USA}
\affiliation{Department of Physics and Astronomy, University of Southern California,
Los Angeles, CA 90089, USA}

\author{Liujun Zou}
\affiliation{Department of Physics, National University of Singapore, 117551, Singapore}

\begin{abstract}
We study the lattice version of higher-form symmetries on tensor-product Hilbert spaces. Interestingly, at low energies, these symmetries may not flow to the topological higher-form symmetries familiar from relativistic quantum field theories, but instead to non-topological higher-form symmetries. We present concrete lattice models exhibiting this phenomenon. One particular model is an $\mathbb{R}$ generalization of the Kitaev honeycomb model featuring an $\mathbb{R}$ lattice 1-form symmetry. We show that its low-energy effective field theory is a gapless, non-relativistic theory with a non-topological $\mathbb{R}$ 1-form symmetry. In both the lattice model and the effective field theory, we demonstrate that the non-topological $\mathbb{R}$ 1-form symmetry is not robust against local perturbations.
In contrast, we also study various modifications of the toric code and their low-energy effective field theories to demonstrate that the compact $\mathbb{Z}_2$ lattice 1-form symmetry does become topological at low energies unless the Hamiltonian is fine-tuned. Along the way, we clarify the rules for constructing low-energy effective field theories in the presence of multiple superselection sectors.
Finally, we argue on general grounds that non-compact higher-form symmetries (such as $\R$ and $\z$ 1-form symmetries) in lattice systems generically remain non-topological at low energies, whereas compact higher-form symmetries (such as $\z_n$ and $U(1)$ 1-form symmetries) generically become topological.

\end{abstract}

\maketitle
\tableofcontents

\section{Introduction}\label{sec:introduction}

Symmetry is a fundamental concept in physics, and in recent years its scope has been extended in various directions, including higher-form symmetries \cite{Gaiotto2015}, non-invertible symmetries \cite{Bhardwaj:2017xup,Chang:2018iay}, multipole symmetries \cite{Gromov:2018nbv}, and modulated symmetries \cite{Pace:2024tgk}, among others. Reviews on generalized global symmetries can be found, for example, in Refs.~\cite{McGreevy_2023,Cordova2022,Brennan:2023mmt,schafernameki2023ictp,Luo:2023ive,Bhardwaj:2023kri,shao2023whats,Carqueville:2023jhb}. In this paper, we focus on higher-form symmetries, which can be defined for both quantum field theories and lattice systems. 

In the context of quantum field theory, we divide higher-form symmetries into two types, \ie topological and non-topological higher-form symmetries. Prototypical examples of topological higher-form symmetries arise in {\it relativistic} quantum field theories, with topological quantum field theories as special cases. In such theories, a $p$-form symmetry with $p\geqslant 0$ is implemented by topological defects supported on codimension-$(p+1)$ submanifolds of spacetime. A defect is topological if 
correlation functions involving it remain invariant under the smooth deformations of its support, provided that during such deformations, the defect does not touch other operators or generate a Dehn twist. In general, a topological $p$-form symmetry acts on operators supported on $p$-dimensional submanifolds in spacetime. Ordinary symmetries acting on local operators are $0$-form symmetries, and higher-form symmetries correspond to the cases with $p\geqslant 1$. When $p\geqslant 1$, the high codimensionality of the symmetry defects allows them to freely pass through one another as long as they are not linked, so these symmetries are Abelian. Moreover, topological higher-form symmetries have no action on local operators, so local perturbations cannot break them.

On the other hand, non-topological higher-form symmetries are also generated by operators supported on codimension-$(p+1)$ submanifolds of spacetime, but they lack the defining topological properties of topological higher-form symmetries. In particular, their symmetry operators supported on spatial submanifolds depend on the detailed shape of those submanifolds, so operators related by smooth deformations need not be equivalent.
As a result, a non-topological higher-form symmetry may act on local operators and can be broken by local perturbations. 
Examples of non-topological higher-form symmetries in quantum field theories can be found in Ref.~\cite{Seiberg:2019vrp}.\footnote{Topological/non-topological higher-form symmetries were referred to as unfaithful/faithful higher-form symmetries in Ref.~\cite{Qi2020} and relativistic/non-relativistic higher-form symmetries in Ref.~\cite{Seiberg:2019vrp}.} Such quantum field theories are necessarily \emph{non-relativistic}.

Higher-form symmetries appear not only in quantum field theories, but also in lattice systems. A particularly important class of systems where higher-form symmetries play a key role are topological orders. As most experimental systems exhibiting evidence of topological orders are in $2+1$ dimensions, where the relevant higher-form symmetries are 1-form symmetries, in this paper we will focus on 1-form symmetries in $(2+1)$-dimensional lattice systems with a tensor-product Hilbert space. However, the general lessons we draw also apply to higher-form symmetries of other form degrees in different dimensions.

How should a 1-form symmetry be defined in such a lattice system? To mimic the 1-form symmetries in field theories described above, these lattice 1-form symmetries should satisfy the following conditions 
\cite{Qi2020, liu2023symmetries, Feng2025}:
\begin{enumerate}[leftmargin=12pt]
    \item Symmetry operators are supported on closed loops, just as 1-form symmetries in continuum field theories.

    \item Symmetry operators on contractible loops mutually commute. In particular, this condition implies that lattice 1-form symmetries must be Abelian, in agreement with the topological 1-form symmetries in continuum field theories.

    \item 
    Symmetry operators can be deformed from one loop to another by the multiplication with symmetry operators on contractible loops. 
    
\end{enumerate}
We will make these notions more precise in the concrete examples below, and regard these conditions as the definition of a 1-form symmetry in a lattice system. We emphasize that Condition 3 does not imply that two symmetry operators that can be deformed into one another are equivalent. In other words, lattice 1-form symmetries are not topological and can therefore act nontrivially on local operators, similar to non-topological 1-form symmetries in non-relativistic quantum field theories. This non-topological property also holds for lattice higher-form symmetries of higher form degrees defined on tensor-factorized Hilbert spaces.

Suppose a lattice system has a higher-form symmetry satisfying the above conditions, does this symmetry flow to a topological or non-topological higher-form symmetry in the low-energy effective field theory? In other words, is there any energy scale below which the lattice higher-form symmetry behaves like a topological or non-topological higher-form symmetry? It is usually expected that such a symmetry flows to a topological higher-form symmetry at low energies. Assuming this expectation is correct, then because local perturbations cannot break a topological higher-form symmetry within the low-energy effective field theory, a higher-form symmetry in a lattice system, even if explicitly broken by certain weak local perturbations, is expected to reemerge at low energies. This expectation is indeed true for many solvable models of topological orders, such as the toric code model, which flows to topological $\mathbb{Z}_2$ gauge theory at low energies. However, in this paper we show that the above expectation can be false, which is exemplified by the models in Sec.~\ref{sec: harmonic oscillator model}, Sec.~\ref{sec:toric_code_no_plaquette}, and Sec.~\ref{app: non-robust toric code}. We further argue that this expectation is generically true if the underlying higher-form symmetry is compact, such as a $\z_n$ or $\U$ 1-form symmetry. If the higher-form symmetry is non-compact, such as an $\R$ or $\z$ 1-form symmetry, then this expectation is false. Our work thus highlights the physical distinctions between lattice higher-form symmetries and topological higher-form symmetries in continuum field theories, and refines the symmetry-guided principle for realizing exotic quantum matter  \cite{Nussinov2006, Nussinov2007}.

We note that the question we address is different from that in Ref. \cite{Liu2025}, which discussed when a quantum state in a lattice system has an emergent higher-form symmetry. In our discussion, whether a lattice higher-form symmetry flows to a topological higher-form symmetry at low energies depends on the Hamiltonian, while the discussion in Ref. \cite{Liu2025} is about a state, without any reference to a Hamiltonian. In Sec. \ref{sec: lattice model with compact Z2 1-form}, we discuss examples of Hamiltonians that can have some common ground states, which are the ground states of the standard toric code model. According to the criterion in Ref. \cite{Liu2025}, such states have a 1-form symmetry. However, as we demonstrate, only for some (but not all) of these Hamiltonians, the lattice 1-form symmetry flows to a topological one at low energies.

The rest of the paper is organized as follows. In Sec.~\ref{sec: harmonic oscillator model}, we present a concrete solvable lattice model with an $\R$ 1-form symmetry, which does not flow to a topological $\R$ 1-form symmetry at low energies. In Sec.~\ref{sec: lattice model with compact Z2 1-form}, we investigate various modifications of the toric code to illustrate that a $\mathbb{Z}_2$ lattice 1-form symmetry generically becomes topological. We also analyze fine-tuned cases where the 1-form symmetry remains non-topological, which in turn motivate a condition for the flow to topological 1-form symmetries, discussed in detail in the next section.
In Sec.~\ref{sec: compactness}, we argue that a compact higher-form symmetry in a lattice system generically flows to a topological higher-form symmetry, whereas a non-compact higher-form symmetry in a lattice system does not. We also clarify the rules for constructing low-energy effective field theories in the presence of multiple superselection sectors. We end this paper with some discussions in Sec.~\ref{sec: discussion}. Various appendices contain further details. In particular, Appendix \ref{app:exact_diagonalization} presents more technical details on the model discussed in Sec.~\ref{sec: harmonic oscillator model}, Appendix \ref{app:EFT} computes the spectrum of the low-energy effective field theory of this model,
and Appendix \ref{app: Z 1-form} discusses lattice systems with a $\z$ 1-form symmetry.

\section{Lattice model with a non-topological $\R$ 1-form symmetry} \label{sec: harmonic oscillator model}

In this section, we introduce an $\R$ generalization of Kitaev's honeycomb model that carries an $\R$ lattice 1-form symmetry, and show that at low energies this symmetry does not flow to a topological 1-form symmetry. In Sec.~\ref{sec: compactness}, we address the general question about when a higher-form symmetry in a lattice system flows to a topological higher-form symmetry at low energies.

\begin{figure}[h!]
    \centering
    \includegraphics[width=0.55\linewidth]{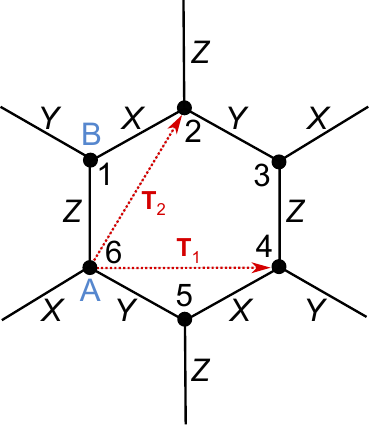}
    \caption{Three different types of links in honeycomb lattice are labelled by $X,Y$ and $Z$. Red arrows $\bT_1$ and $\bT_2$
    are lattice vectors. Two sublattices are labelled by $A$ and $B$.
    }
    \label{fig:label_of_links}
\end{figure}

The model is defined on a two dimensional honeycomb lattice, and at each site of the lattice there is an oscillator (see Fig.~\ref{fig:label_of_links}).{\footnote{Here we slightly abuse the terminology: we mean only that the Hilbert space at each site is isomorphic to that of a harmonic oscillator, whereas the Hamiltonian need not take a harmonic-oscillator form.}} We will view each hexagon of the lattice as an elementary loop, and we assume that on each of these loops there is a symmetry generator:
\begin{equation} \label{eq: generators}
Q_{l}=\alpha(p_{1}-p_{2}+p_{4}-p_{5})+\beta(x_{1}-x_{3}+x_{4}-x_{6}),
\end{equation}
where $x_i$ and $p_i$ are the position and momentum operators of the oscillator at site $i$ in Fig.~\ref{fig:label_of_links}, respectively, and $\alpha$ and $\beta$ are two non-zero real numbers. Here the index of the site is defined locally for each hexagon. For example, site 1 of the middle hexagon in Fig.~\ref{fig:label_of_links} is identified with site 5 of the upper left hexagon. Such a generator generates an $\mathbb{R}$-parametrized symmetry
\beq
W_l(t)=e^{iQ_lt}, 
\eeq
where $t\in\R$.

\begin{figure}
    \centering
    \includegraphics[width=\linewidth]{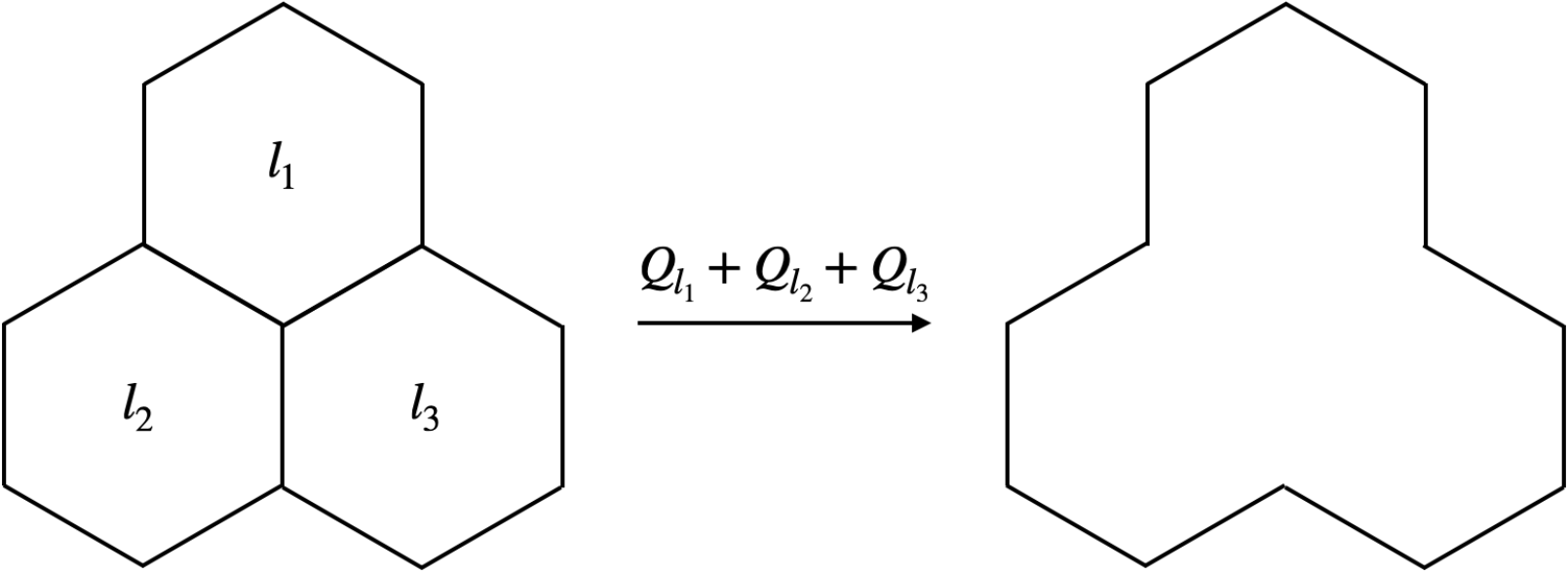}
    \caption{When the $Q_l$ operators from 3 adjacent elementary loops (\ie hexagons) are added together, the operators in the interior all cancel, and the result is an operator supported entirely in the exterior, which can be viewed as a larger loop. It is this condition that allows us to view the supports of the symmetry operators as closed loops.}
    \label{fig: loop addition}
\end{figure}

One can check that the symmetry defined above satisfies the definitions of a lattice 1-form symmetry, discussed in the Introduction. Concretely, in addition to being supported on closed loops, the generators $Q_l$ satisfy two important properties:
\begin{enumerate}[leftmargin=12pt]

    \item When multiple $Q_l$'s from adjacent hexagons are added together, the support of the resulting operator lies solely on the boundary of the larger loop, with no support in its interior (see Fig.~\ref{fig: loop addition} for an example involving three adjacent hexagons).

    \item $[Q_{l_1}, Q_{l_2}]=0$ for any pair of elementary loops (\ie hexagons) $l_1$ and $l_2$.
    
\end{enumerate}
This suggests that the symmetry generated by $Q_l$ should be interpreted as a lattice version of $\R$ 1-form symmetry.

\begin{figure}
    \centering
    \includegraphics[width=0.7\linewidth]{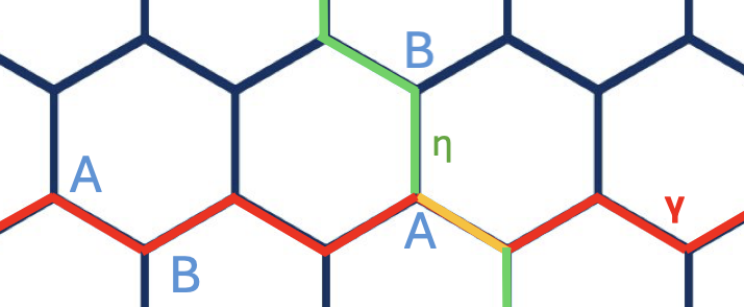}
    \caption{Non-contractible loops on the honeycomb lattice, denoted by $\gamma$ (red loop) and $\eta$ (green loop), respectively.}
    \label{fig: non-contractible loops}
\end{figure}

We remark that if the Hamiltonian is local and the system is put on a torus with periodic boundary conditions along both directions, then the symmetries generated by $Q_l$ in Eq.~\eqref{eq: generators} imply the existence of symmetry generators supported on the two non-contractible loops of the torus (denoted by $\eta$ and $\gamma$, see Fig.~\ref{fig: non-contractible loops}):
\beq \label{eq: non-contractible loops}
\begin{split}
Q_\eta &=\sum_{i\in A_\eta}x_i-\sum_{i\in B_\eta}x_i,\\
Q_\gamma &=\sum_{i\in A_\gamma}p_i-\sum_{i\in B_\gamma}p_i,
\end{split}
\eeq
where $i\in A_\eta$ (resp.~$i\in B_\eta$) means that the site $i$ is in the $A$ (resp.~$B$) sublattice on the non-contractible loop $\eta$, and similar for the loop $\gamma$. These non-contractible symmetry generators, together with the contractible generators $Q_l$ on elementary loops, form an algebra:
\beq \label{eq: loop commutators}
\begin{split}
[Q_\eta, Q_l]&=[Q_\gamma, Q_l]=0,\\
[Q_\eta, Q_\gamma]&=2i,\\
e^{iQ_\eta t_\eta}e^{iQ_\gamma t_\gamma}&=e^{-2i t_\eta t_\gamma}e^{iQ_\gamma t_\gamma}e^{iQ_\eta t_\eta},
\end{split}
\eeq
where $t_{\eta,\gamma}\in\R$. The last relation implies that all energy levels of any Hamiltonian with this $\R$ 1-form symmetry must be infinitely degenerate on a torus, and these degenerate states are related by the actions of the non-contractible symmetry operators $e^{iQ_\eta t_\eta}$ and $e^{iQ_\gamma t_\gamma}$.\footnote{By deforming the non-contractible 1-form symmetry operators using the contractible ones, it is straightforward to see that these degenerate states are locally indistinguishable. Namely, for any local operator $O$ and any two orthogonal degenerate states $|m\rangle$ and $|n\rangle$, $\langle m|O|n\rangle=C_O\delta_{mn}$, where $C_O$ is a constant that depends on $O$.} It indicates an anomaly in the $\mathbb{R}$ lattice 1-form symmetry.

A class of Hamiltonians with such an $\R$ 1-form symmetry is the Kitaev-type Hamiltonian:
\beq\label{eq:free_boson_Hamiltonian}
H=\sum_{\mu=X,Y,Z}\sum_{\langle i,j\rangle\in\mu}J_{\mu}f_{\mu}(x_{i},p_{i},x_{j},p_{j}).
\eeq
Here all bonds in the honeycomb lattice are divided into three types, \ie $X$, $Y$ and $Z$ bonds (see Fig.~\ref{fig:label_of_links}), and $\langle i, j\rangle\in\mu$ means that two adjacent sites $i$ and $j$ are connected by a $\mu$-type bond \cite{Kitaev2006}. The parameters $J_\mu>0$, the function $f_X$ is any real function of $x_i+x_j$ that is bounded from below, $f_Y$ is any real function of $\alpha p_i+\beta x_i+\alpha p_j+\beta x_j$ that is bounded from below, and $f_Z$ is any real function of $p_i+p_j$ that is bounded from below. For simplicity, we can take
\beq\label{eq:local_terms_in_Hamiltonian}
\begin{split}
    f_{X}&=(x_{i}+x_{j})^{2},\\
    f_{Y}&=(\alpha p_{i}+\beta x_{i}+\alpha p_{j}+\beta x_{j})^{2},\\
    f_{Z}&=(p_{i}+p_{j})^{2}.
\end{split}
\eeq
With this choice, the Hamiltonian describes non-interacting bosons. We will refer to this particular Hamiltonian as the $\mathbb{R}$ Kitaev model.

Below, we study this $\mathbb{R}$ Kitaev model on a torus with periodic boundary conditions. We will see that all energy levels are indeed infinitely degenerate, enforced by the $\R$ lattice 1-form symmetry. However, we will also show that after adding a weak local perturbation, the resulting model has a unique gapped ground state. This implies that the $\R$ lattice 1-form symmetry does not flow to a topological $\R$ 1-form symmetry, as it does not reemerge at low energies.

\subsection{Degenerate energy spectrum and its {non-robustness}} \label{subsec: free boson spectrum}

As we will soon see, it is beneficial to count the number of independent conserved quantities for the Hamiltonian defined in Eqs.~\eqref{eq:free_boson_Hamiltonian} and \eqref{eq:local_terms_in_Hamiltonian}. Below, we only consider conserved quantities that are linear in the $x$'s and $p$'s, like in Eq.~(\ref
{eq: generators}). Each elementary loop contributes one conserved quantity, $Q_l$, so seemingly we have $L_{1}L_{2}$ conserved quantities, if the system is on a torus with size $L_1\times L_2$, where $L_{1,2}$ is the number of unit cells along the two translation vectors. However, these conserved quantities are not independent, since $\sum_lQ_l=0$ under periodic boundary conditions. After taking this constraint into account, we only get $L_{1}L_{2}-1$ independent conserved quantities, and they all commute with each other. Additionally, there are two more conserved quantities associated with the two non-contractible loops, $Q_\eta$ and $Q_\gamma$ in Eq.~\eqref{eq: non-contractible loops}, which have commutation relations in Eq.~\eqref{eq: loop commutators}. Therefore, the total number of independent conserved charges related to the $\mathbb{R}$ lattice 1-form symmetry is $L_{1}L_{2}+1$.

According to Refs.~\cite{COLPA1986377, COLPA1986417}, the quadratic Hamiltonian as in Eqs.~\eqref{eq:free_boson_Hamiltonian} and \eqref{eq:local_terms_in_Hamiltonian} can be brought into the following standard form by a bosonic Bogoliubov transformation:
\beq\label{eq:normal_modes}
H=\sum_{a=0}^{2L_{1}L_{2}-1}(u_{a}\tilde{x}_{a}^{2}+v_{a}\tilde{p}_{a}^{2}),
\eeq
where $a$ indexes the normal modes, $u_{a}, v_{a}\geqslant 0$ are constants (which are non-negative as $H$ is bounded from below), and we sum over all normal modes $\tilde{x}_{a}$'s and $\tilde{p}_{a}$'s, which are linear combinations of $x_{i}$'s and $p_{i}$'s and satisfy $[\tilde x_{a_1}, \tilde x_{a_2}]=[\tilde p_{a_1}, \tilde p_{a_2}]=0$ and $[\tilde x_{a_1}, \tilde p_{a_2}]=i\delta_{a_1, a_2}$. The spectrum of the $\mathbb{R}$ Kitaev model is determined analytically in Appendix \ref{app:exact_diagonalization}, but to gain physical insights, we do not need to perform the diagonalization or to determine the values of $u_{a}$ and $v_{a}$ explicitly. Notice that in general the normal modes can be classified into the following 3 types:

\begin{enumerate}[label=(\Roman*)]
    
    \item If $u_{a},\,v_{a}\not=0$, this mode behaves as a harmonic oscillator and contributes a discrete energy spectrum. It does not contribute any conserved quantity linear in $x$'s and $p$'s.
    
    \item 
    If either $u_{a}=0$, $v_a\neq 0$ or $v_{a}=0$, $u_a\neq0$, this mode behaves as a free particle and contributes a continuous 
    energy spectrum. Each such mode contributes 1 conserved quantity linear in $x$'s and $p$'s. If $u_{a}=0$, then $\tilde{p}_{a}$ is conserved, and if $v_a=0$, then $\tilde{x}_a$ is conserved. Without loss of generality, below we assume $u_{a}=0$ for this type.
    
    \item 
    If $u_{a}=v_{a}=0$, this mode is associated with one (zero-energy) level of infinite degeneracy. Both $\tilde{x}_{a}$ and $\tilde{p}_{a}$ are conserved quantities, so each such zero mode contributes 2 conserved quantities linear in $x$'s and $p$'s, \ie $\tilde x_a$ and $\tilde p_a$. Importantly, these two conserved quantities do not commute and $[\tilde{x}_{a},\tilde{p}_{a}]=i$.
    
\end{enumerate}

To recover the number of conserved quantities due to the $\mathbb{R}$ lattice 1-form symmetry and their commutation relations in Eq.~\eqref{eq: loop commutators}, the system must have at least $L_{1}L_{2}-1$ type II modes and 1 type III mode. Having additional type II or type III modes would introduce conserved quantities not due to the lattice 1-form symmetry, which can be viewed as a consequence of fine tuning. In Appendix \ref{app:exact_diagonalization}, we confirm that this model has exactly $L_1L_2-1$ type II modes and 1 type III mode. The reason is that we find a normal mode for each momentum $\bk$ with a level spacing
\begin{equation}\label{eq:type I level-spacing}
\begin{split}
    &E_1 (\bk) = 2\sqrt{2 J_x J_y J_z}\\
    & \sqrt{ \frac{\beta^2}{J_x} \cos^2\! \frac{\bk\!\cdot\!\bT_{12}}{2}\!+\!\frac{\alpha^2}{J_z} \cos^2\!\frac{\bk\!\cdot\!\bT_1}{2}\!+\!\frac{1}{J_y} \cos^2\! \frac{\bk\!\cdot\!\bT_2}{2} },
\end{split}
\end{equation}
where $\bT_{12}\equiv \bT_1-\bT_2$. Clearly, $E_1(\bk)$ does not vanish at any $\bk$ as long as $J_{x,y,z}> 0$ and $\alpha,\beta\neq 0$, so these modes are of type I. Since the model has in total $2L_1L_2$ normal modes, excluding these $L_1L_2$ type I modes, the remaining $L_1L_2$ normal modes must all be of type II or type III, and there must be exactly $L_1L_2-1$ type II modes and 1 type III mode.

Here we arrange these normal modes so that $a=0$ labels the unique type III mode and $a=1,2,\dots,L_{1}L_{2}-1$ label the type II modes. From the commutation relation and lattice symmetries, the type III mode should be given by{\footnote{To see it, notice that translation symmetry requires these operators have a definite momentum. Moreover, for generic values of $\alpha$, $\beta$ and $J_{X,Y,Z}$, the model has an inversion symmetry, which implies that the momentum of the unique type III mode must be one of $(0, 0)$, $(0, \pi)$, $(\pi, 0)$ and $(\pi, \pi)$ in the Brillouin zone. Furthermore, for some specific values of $\alpha$, $\beta$ and $J_{X,Y,Z}$, the model has an additional 3-fold rotational symmetry, which implies that the momentum of the type III mode must be $(0, 0)$ in this case. Because changing the values of $\alpha$, $\beta$ and $J_{X,Y,Z}$ is not expected to change the momentum of the type III mode discontinuously, the momentum of this mode should be $(0, 0)$ for generic values of $\alpha$, $\beta$ and $J_{X,Y,Z}$. Knowing that this type III mode involves the non-contractible loop operators and has zero momentum, the expressions of $\tilde x_0$ and $\tilde p_0$ then follow.}}
\beq\label{eq:typeIII}
\begin{split}
    \tilde{x}_{0}=\frac{1}{\sqrt{2L_1 L_2}}\sum_\eta Q_{\eta} \propto \sum_{i\in A} x_i - \sum_{i\in B} x_i,  \\
    \tilde{p}_{0}=\frac{1}{\sqrt{2L_1 L_2}}\sum_\gamma Q_{\gamma} \propto \sum_{i\in A} p_i - \sum_{i\in B} p_i,
\end{split}
\eeq
hence the type III mode is actually connected to the non-contractible loop operators in Eq.~(\ref{eq: non-contractible loops}). Similarly, conserved momenta $\tilde{p}_{a}$ of the $L_1L_2-1$ type II modes are related to the generators in Eq.~(\ref{eq: generators}) by an invertible linear map
\beq
\tilde{p}_{a}=\sum_{l}A_{a,l}Q_{l},
\eeq
with $A_{a,l}$ an $(L_{1}L_{2}-1)\times (L_{1}L_{2}-1)$ invertible matrix.

Due to the anomaly of the $\R$ lattice 1-form symmetry, the ground states are uncountably infinitely degenerate (for any system size).{\footnote{ The degenerate ground states can be non-normalizable.}}
If these ground states are taken as the eigenstates of $\tilde p_0$, then different ground states are related by applying $\tilde x_0$. This also means that the 1-form symmetry is spontaneously broken.

However, this infinite degeneracy is by no means robust. We can turn all the modes into type I by adding suitable weak local perturbations that explicitly break the $\R$ lattice 1-form symmetry. For example, we can add $\sum_iV_i$ with $V_{i}=\epsilon (x_{i}^{2}+p_{i}^{2})$ for each site $i$ (where $\epsilon$ is a small positive parameter) to Eq.~(\ref{eq:free_boson_Hamiltonian}) and then repeat the above symmetry analysis. We find no conserved quantity linear in $x$ and $p$ this time, implying that all modes are of type I. The ground state is therefore non-degenerate on the torus, regardless of its size. In fact, solving the spectrum as explained in Appendix \ref{app:exact_diagonalization}, we find two type I modes for each momentum, one with a level spacing $E_1(\bk)$ in Eq.~\eqref{eq:type I level-spacing} (to leading order in $\epsilon$) and another with a level spacing $E_2(\bk)$ satisfying
\begin{equation}
\begin{split}
     &\frac{E^2_1 (\bk) E^2_2 (\bk)}{4J_xJ_yJ_z} =
     2\epsilon \Big[\alpha^2\sin^2\!{\frac{\bk\!\cdot\!\bT_{12}}{2}}\!+\!\beta^2 \sin^2\!{\frac{\bk\!\cdot\!\bT_{1}}{2}}\Big]+ \\
     &\epsilon^2\Bigg[\frac{1}{J_y}\!+\!\frac{\alpha^2\!+\!\beta^2\sin^2\!{\dfrac{\bk\!\cdot\!\bT_{1}}{2}}}{J_z}\!+\!\frac{\beta^2\!+\!\alpha^2\sin^2\!{\dfrac{\bk\!\cdot\!\bT_{12}}{2}}}{J_x}\Bigg] 
     \!+\!\mathcal{O}(\epsilon^3).
\end{split}
\end{equation}
The level spacing $E_2(\bk)$ is non-zero for every $\mathbf{k}$. It is of order  $\mathcal{O}(\sqrt{\epsilon})$ at $\mathbf{k}\neq 0$ and $\mathcal{O}(\epsilon)$ at $\mathbf{k}= 0$.
Thus, the energy gap above the unique gapped ground state is of order $\mathcal{O}(\epsilon)$. In other words, the anomalous $\mathbb{R}$ 1-form symmetry, which would imply an infinite level degeneracy, does not reemerge at low energies if it is broken explicitly by a weak local perturbation. The above analysis shows that the $\R$ lattice 1-form symmetry in this model does not flow to a topological $\R$ 1-form symmetry at low energies.

\subsection{Effective field theory} \label{subsec: effective field theory}

The above discussion of the $\mathbb{R}$ 1-form symmetry suggests that the low-energy effective field theory of the $\mathbb{R}$ Kitaev model is an exotic gapless non-relativistic quantum field theory. 
We propose that it is captured by the following Lagrangian
\begin{equation} \label{eq:EFT_Lagrangian}
\mathcal{L}=A_y\partial_t A_x -J(\partial_x A_y-\partial_y A_x)^2,
\end{equation}
where $A_x,A_y$ are $\mathbb{R}$-valued fields. We emphasize that although $A_x,A_y$ resemble components of a gauge field, the theory does not have any gauge symmetry.

The equations of motion of the effective field theory can be recast into current conservation equations
\begin{equation}
\begin{aligned}
    &\partial_t j^{tx}+\partial_y j^{yx}=0,
    \\
    &\partial_t j^{ty}+\partial_x j^{xy}=0,
    \end{aligned}
\end{equation}

for an antisymmetric current $j^{\mu\nu}$ with components
\begin{equation}
    j^{tx}=A_y,\ \, j^{ty}=-A_x,\ \, j^{xy}=-2J(\partial_x A_y-\partial_y A_x).\label{eq:current_component}
\end{equation}
These current conservation equations resemble those of an $\mathbb{R}$ topological 1-form symmetry in 2+1 dimensions, except that the condition $B\equiv \partial_x j^{tx}+\partial_y j^{ty}=0$ is missing. As we will show below, they instead describe a non-topological $\mathbb{R}$ 1-form symmetry.

The non-zero operator $B$ is itself a locally conserved quantity obeying $\partial_t B=0$, which can be interpreted as the low-energy version of the $Q_l$ operator of the $\mathbb{R}$ lattice 1-form symmetry in Eq.~\eqref{eq: generators}.
Integrating it over an open region $M$ gives a conserved loop operator supported on the boundary $\partial M$:
\begin{equation}
\begin{aligned}
    Q(\partial M)=\int_M B&=\oint_{\partial M} (j^{tx}\, dy-j^{ty }\, dx).
    \end{aligned}
\end{equation}
Since these contractible loop operators are all integrals of $B$, they mutually commute as 1-form symmetry operators. However, unlike in the case of topological 1-form symmetries, they are non-zero operators depending on the details of $\partial M$. Thus, they generate a non-topological $\mathbb{R}$ 1-form symmetry, closely paralleling the $\mathbb{R}$ lattice 1-form symmetry in lattice systems.

On a torus, this non-topological $\mathbb{R}$ 1-form symmetry has additional conserved charges wrapping around the non-contractible cycles,
\begin{equation}
\begin{aligned}
    Q_x=\oint j^{ty}\, dx,
    \quad
    Q_y=\oint j^{tx}\, dy,
\end{aligned}
\end{equation}
matching the non-contractible charges of the $\mathbb{R}$ lattice 1-form symmetry in Eq.~\eqref{eq: non-contractible loops}.

Since $j^{ty}=-A_x$ and $j^{tx}=A_y$ are conjugate variables, upon quantization, these two non-contractible charges do not commute: 
\begin{equation}
    [j^{ty}(\mathbf{x}),j^{tx}(\mathbf{y})]=-i\delta^2(\mathbf{x}-\mathbf{y})\ \Rightarrow\ [Q_x, Q_y]=-i.
\end{equation}
This reproduces the nontrivial commutation relation on the lattice in Eq.~\eqref{eq: loop commutators} that enforces an infinite degeneracy at every energy level.

As shown in Appendix \ref{app:EFT}, this effective field theory also reproduces the low-energy spectrum of the $\mathbb{R}$ Kitaev model, which has a type II mode at every non-zero momentum $\mathbf{k}\neq0$ and a type III mode at $\mathbf{k}=0$.

Physically, this theory can be interpreted as coupling a Chern-Simons theory with gauge group $\mathbb{R}$ to a special matter system with local conservation laws, described by the following Lagrangian
\begin{equation}
    \begin{aligned}
&\mathcal{L}=\mathcal{L}_{\text{CS}}+\mathcal{L}_{\text{M}}  ,
  \\
  &\mathcal{L}_{\text{CS}}=A_y\partial_t A_x+ A_t(\partial_x A_y-\partial_y A_x),
  \\
  &\mathcal{L}_{\text{M}}=\frac{1}{4J}(\partial_t\phi-A_t)^2,\label{eq:CS+M}
  \end{aligned}
\end{equation}
where $A_\mu$ is an $\mathbb{R}$ gauge field and $\phi$ is its Stueckelberg field that transforms under the $\mathbb{R}$ gauge symmetry,
\begin{equation}
    A_\mu\rightarrow A_\mu+\partial_\mu\alpha,\quad \phi\rightarrow\phi+\alpha.\label{eq:R_gauge_symmetry}
\end{equation}
The $\mathbb{R}$ Chern-Simons theory, when not coupled to the matter, has an $\mathbb{R}$ topological 1-form symmetry generated by the topological Wilson loops $W_t=\exp(i t \oint A)$, $t\in \mathbb{R}$. The topological property of the Wilson loops follows from the equations of motion, which we recast into the form of current conservation equations:\footnote{Strictly speaking, the current $j^{\mu\nu}=\epsilon^{\mu\nu\sigma}A_\sigma$ is not gauge invariant. Thus, the $\mathbb{R}$ 1-form symmetry of the $\mathbb{R}$ Chern-Simons theory is an example of continuous global symmetry without Noether currents. See Ref.~\cite{Harlow:2018tng} for other such examples.}
\begin{equation}\label{eq:R_current_conservation}
    \partial_\mu j^{\mu\nu}=0,\quad j^{\mu\nu}=\epsilon^{\mu\nu\sigma}A_\sigma.
\end{equation}
The matter system is a collection of decoupled oscillators, one for each point, and each of them has an independent conserved quantity
\begin{equation}
\partial_t j_t=0,\quad j_t=\partial_t\phi.    
\end{equation}
Coupling the matter system to the $\R$ Chern-Simons theory Higgses the $\mathbb{R}$ gauge symmetry in Eq.~\eqref{eq:R_gauge_symmetry} and turns the $\mathbb{R}$ gauge field to an $\mathbb{R}$-valued vector field as in Eq.~\eqref{eq:EFT_Lagrangian}.
After gauge fixing $\phi=0$ and integrating out $A_t$, we recover the Lagrangian of the effective field theory in Eq.~\eqref{eq:EFT_Lagrangian}.
Since the matter theory has only charge density but no local current, coupling it to the $\mathbb{R}$ Chern-Simons theory relaxes only the Gauss law, 
\begin{equation}
    \begin{aligned}
    \partial_x j^{tx}+\partial_y j^{ty}= \frac{1}{2J}(\partial_t\phi-A_t)\neq0,
    \end{aligned}
\end{equation}
and keeps intact the other current conservation equations in Eq.~\eqref{eq:R_current_conservation}. This has the effect of turning the topological $\mathbb{R}$ 1-form symmetry to a non-topological 1-form symmetry.
This mechanism is similar to the one used in Ref.~\cite{Seiberg:2019vrp} to construct non-relativistic field theories with non-topological 1-form symmetries.

Lastly, we comment on the robustness of the effective field theory. The theory can be gapped out to a unique gapped ground state by adding a weak local perturbation $\delta\mathcal{L}=\epsilon[(A_x-\partial_x\phi)^2+(A_y-\partial_y\phi)^2]$ that explicitly breaks the non-topological $\mathbb{R}$ 1-form symmetry. This means that this theory is not robust, in stark contrast to a pure $\mathbb{R}$ Chern-Simons theory, which has no gauge-invariant local operators and is therefore manifestly robust. This lack of robustness of the non-topological $\mathbb{R}$ 1-form symmetry is consistent with the non-robustness of the $\mathbb{R}$ lattice 1-form symmetry in the underlying microscopic lattice model.

\section{Lattice model with a $\mathbb{Z}_2$ lattice 1-form symmetry}\label{sec: lattice model with compact Z2 1-form}

To develop a holistic understanding of lattice higher-form symmetries, and to contrast with the above model with a non-compact 1-form symmetry, we now turn to models with a compact 1-form symmetry. In this section, we discuss lattice models with a $\mathbb{Z}_2\times\mathbb{Z}_2$ lattice 1-form symmetry, including the  toric code model \cite{Kitaev2003} and its various modifications \cite{Moessner_2001,Bravyi2010}. We construct the low-energy effective field theories for these models and analyze how the $\mathbb{Z}_2\times\mathbb{Z}_2$ lattice 1-form symmetry is realized in these low-energy effective field theories. Along the way, we clarify the rules for constructing effective field theory in systems with multiple superselection sectors, which eventually leads to a necessary condition for topological higher-form symmetry to be formulated in the next section.

\subsection{Toric code}
We begin by reviewing the toric code and its low-energy effective theory. 
Readers familiar with this material may skip this subsection.

The toric code is defined on a square lattice with a qubit placed on each link \cite{Kitaev2003}. On link $i$, there is a pair of Pauli operators,  $X_i$ and $Z_i$, acting on the qubit. The Hamiltonian of the toric code is given by
\beq \label{eq:toric_code}
H=-J_1\sum_vA_v-J_2\sum_{ p}B_{p},
\eeq
with $J_{1,2}>0$. Here $A_v=\prod_{i\in v}X_i$ is a vertex term with $v$ labeling the vertices of the lattice and the product running over links $i\in v$ connected to the vertex $v$. $B_p=\prod_{i\in p}Z_i$ is a plaquette term with $p$ labeling the plaquettes of the lattice and the product running over links $i\in p$ forming the plaquette $p$.

\begin{figure}[h!]
    \centering
    \includegraphics[width=0.7\columnwidth]{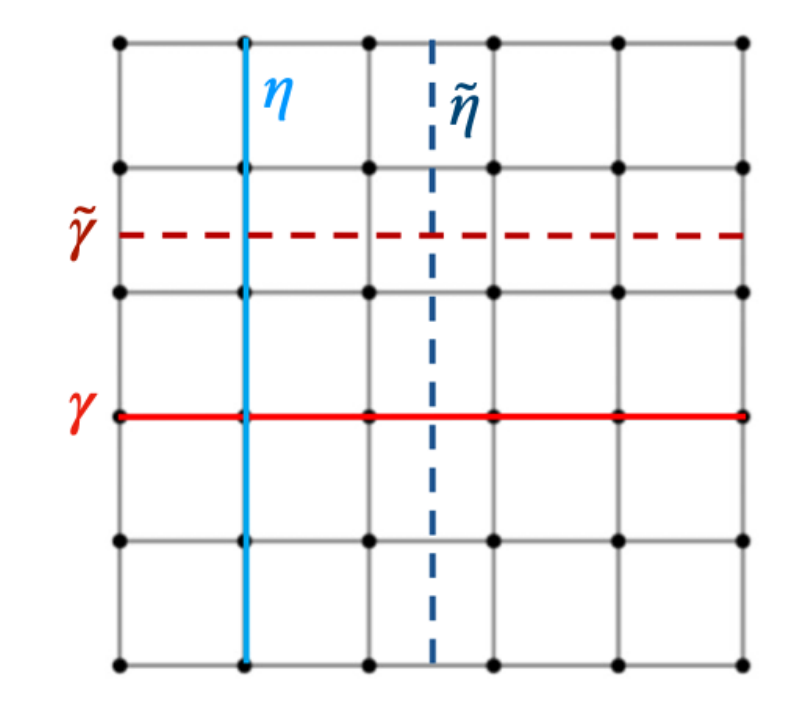}
    \caption{Non-contractible loop operators of toric code supported on the lattice and the dual lattice.}
    \label{fig:loop_on_lattice_and_dual_lattice}
\end{figure}

This lattice model has a $\mathbb{Z}_2^{(a)}\times\mathbb{Z}_2^{(b)}$ 1-form symmetry generated by the vertex $A_v$ and plaquette $B_p$ operators, which are supported on closed loops on the lattice and the dual lattice, respectively. It is easy to check that for every vertex $v,v'$ and plaquette $p,p'$,
\begin{equation}
    \begin{aligned}
        &[A_v, H]=[B_p, H]=0,
        \\
        &[A_{v},A_{v'}]=[A_{v},B_{p}]=[B_{p},B_{p'}]=0,
    \end{aligned}
\end{equation}
and the multiplication of adjacent small loop operators becomes symmetry operators supported on larger loops, hence all three conditions for lattice 1-form symmetries as laid out in the Introduction are satisfied. Notably, the $\mathbb{Z}_2^{(a)}\times\mathbb{Z}_2^{(b)}$ 1-form symmetry has a mixed 't Hooft anomaly that can be detected by the non-contractible loop operators when the system is put on a torus
\beq\label{eq:Wilson_loop_and_'tHooft_loop_toric_code}
\begin{split} W_{\gamma}&=\prod_{i\in\gamma}Z_{i}\;,\quad W_{\eta}=\prod_{i\in\eta}Z_{i}\;,\\
T_{\tilde{\gamma}}&=\prod_{i\in\tilde{\gamma}}X_i\;,\quad T_{\tilde{\eta}}=\prod_{i\in\tilde{\eta}}X_i\;,
\end{split}
\eeq
where $i\in\gamma,\eta$ means that the link $i$ belongs to loop $\gamma,\eta$, and $i\in\tilde\gamma,\tilde\eta$ means that the link $i$ intersects with loop $\tilde\gamma,\tilde\eta$ (see Fig.~\ref{fig:loop_on_lattice_and_dual_lattice}). These non-contractible operators obey the following non-commutative algebra
\beq \label{eq: toric code commutator}
    W_{\gamma}T_{\tilde{\eta}}&=-T_{\tilde{\eta}}W_{\gamma},\quad W_{\eta}T_{\tilde{\gamma}}&=-T_{\tilde{\gamma}}W_{\eta},    
\eeq
which indicates a mixed anomaly and further implies a four-fold degeneracy at every energy level.

The toric code has a gapped spectrum with ground states satisfying
\begin{equation}\label{eq: TC ground state A and B}
A_v = 1 , \quad B_p = 1
\end{equation}
for every vertex $v$ and plaquette $p$. 
It means that the $\mathbb{Z}_2^{(a)} \times \mathbb{Z}_2^{(b)}$ lattice 1-form symmetry becomes topological in the ground state subspace and hence flows to a topological $\mathbb{Z}_2^{(a)} \times \mathbb{Z}_2^{(b)}$ 1-form symmetry in the low-energy effective field theory.

At low energies, the toric code is described by the effective Lagrangian of $\z_2$ gauge theory \cite{Zee1995, wen2004quantum}

\beq
\mc{L}_{\mathbb{Z}_2}=\frac{2}{2\pi}\epsilon^{\mu\nu\lambda}a_\mu \partial_\nu b_\lambda,\label{eq:Z2_gauge_theory}
\eeq
where $b_\mu$ and $a_\mu$ are $U(1)$ 1-form gauge fields. Their equations of motion, respectively  
\begin{equation}
\partial_\mu a_\nu-\partial_\nu a_\mu=0,\quad\partial_\mu b_\nu-\partial_\nu b_\mu=0,
\label{eq: curl a and b vanish}
\end{equation}
implies that the Wilson loops
\begin{equation}
    \mathcal{W}_C = \exp\left(i\oint_C a\right),\quad \mathcal{T}_C = \exp\left(i\oint_C b\right),
\end{equation}
are topological operators. Furthermore, because $U(1)$ gauge fields can have non-trivial quantized fluxes through closed 2-manifolds, after summing over flux sectors $\oint da,\, \oint db\in 2\pi\mathbb{Z}$, the holonomy are constrained such that  $\mathcal{W}_C,\, \mathcal{T}_C=\pm1$.\footnote{To see this in more detail, consider the spacetime manifold as $\mathcal{M}=T^2 \times S^1$ (where $T^2=S^1\times S^1$ is the spatial torus and $S^1$ is the temporal direction). The exponentiated action takes the form $\exp(i\oint_\mathcal{M} \mathcal{L}_{\mathbb{Z}_2}) = \exp \left[\left(\frac{i}{\pi} \oint_{S^1}a\right)\left(\oint_{S^1\times S^1}db\right)\right]$, where flux-quantization restricts $\oint_{S^1\times S^1}db\in 2\pi\mathbb{Z}$. Hence, upon summing over flux sectors in the partition function, only $\oint a \in \pi \mathbb{Z}$ contributes. Similarly, $\oint b \in \pi \mathbb{Z}$.} 
Thus, these Wilson loops generate a \textit{topological} $\mathbb{Z}_2^{(a)}\times\mathbb{Z}_2^{(b)}$ 1-form symmetry, which is the low-energy image of the $\mathbb{Z}_2^{(a)}\times\mathbb{Z}_2^{(b)}$ lattice 1-form symmetry discussed above. For a contractible loop $C$, Eq.~\eqref{eq: curl a and b vanish} implies $\mathcal{W}_C=\mathcal{T}_C=1$, which corresponds to Eq.~\eqref{eq: TC ground state A and B} on the lattice. On a torus, the Wilson loops $\mathcal{W}_{x,y}, \mathcal{T}_{x,y}$ wrapping around non-contractible cycles in the $x,y$ direction reproduce an algebra identical to Eq.~\eqref{eq: toric code commutator}
\begin{equation}
    \mathcal{W}_x \mathcal{T}_y=  -\mathcal{T}_y \mathcal{W}_x,\quad \mathcal{W}_y \mathcal{T}_x=  -\mathcal{T}_x \mathcal{W}_y,
\end{equation}
which again implies a four-fold degeneracy. Importantly, the topological properties of these Wilson loops implies that there are no local operators charged under this 1-form symmetry, and thus the lattice 1-form symmetry is robust against any weak local perturbations. It in particular further implies the robustness of the four-fold ground-state degeneracy in the thermodynamic limit. 

\subsection{Odd toric code}
\label{sec:odd_toric_code}

We now discuss the first modification of the toric code Hamiltonian in Eq.~\eqref{eq:toric_code}, where we take $J_2<0$ while keeping $J_1>0$. This model is known as the odd toric code model \cite{Moessner_2001} because it has a similar gapped spectrum as the toric code except that its ground states satisfy
\begin{equation}
    A_v=1,\quad B_p=-1,
\end{equation}
for all vertices $v$ and all plaquettes $p$, with $B_p$ taking the opposite eigenvalue.
Here, we assume that $L_x L_y$ is even, so that $\prod_p B_p=1$ is satisfied. 

This Hamiltonian preserves the $\mathbb{Z}_2^{(a)}\times\mathbb{Z}_2^{(b)}$ lattice 1-form symmetry generated by $A_v$ and $B_p$, respectively. The commutation relation in Eq.~\eqref{eq: toric code commutator} again implies a four-fold ground state degeneracy. In the ground state subspace, the $\mathbb{Z}_2^{(a)}$ 1-form symmetry becomes topological, while the $\mathbb{Z}_2^{(b)}$ 1-form symmetry appears non-topological since $B_p=-1$, implying that the expectation value of a contractible symmetry operator depends on the number of plaquettes it encloses, contradictory to the topological property of a topological 1-form symmetry discussed in the Introduction. This non-topologicalness is however innocuous since it can be fixed by redefining the symmetry operator on each plaquette by a $c$-number from $B_p$ to $\tilde B_p=-B_p$. After this redefinition, the new symmetry operators are topological in the ground state subspace. Hence, we conclude that the $\mathbb{Z}_2^{(a)}\times\mathbb{Z}_2^{(b)}$ 1-form symmetry becomes topological in the ground state subspace.

In fact, when $L_xL_y$ is even, the odd toric code Hamiltonian is related to the original toric code Hamiltonian by a unitary transformation. Without the lost of generality, we assume $L_x$ is even. In this case, we can relate the two Hamiltonians by the unitary transformation 
\beq
U=\prod_{i=1}^{L_x/2} \prod_{j=1}^{L_y} X^{(y)}_{2i,j}.
\eeq
Here, $X^{(y)}_{i,j}$ denotes the $X$ operator on the $+y$-direction-oriented link from site $(i,j)$. Under this unitary transformation, $B_p$ is mapped to $-B_p$ and the redefined $\mathbb{Z}_2^{(b)}$ 1-form symmetry operators $\tilde B_p$ in the odd toric code is mapped to the $\mathbb{Z}_2^{(b)}$ 1-form symmetry operators $B_p$ in the original toric code.

At low energies, the odd toric code is described by the following effective Lagrangian
\beq
\mathcal{L}=\mc{L}_{\mathbb{Z}_2}-\frac{1}{\ell^2}a_t,\label{eq:Z2_gauge_theory_mesh}
\eeq
where $\mathcal{L}_{\mathbb{Z}_2}$ is the $\mathbb{Z}_2$ gauge theory Lagrangian in Eq.~\eqref{eq:Z2_gauge_theory} and $\ell$ is a characteristic length scale such that $\ell^2$ is the area of a lattice unit cell. The second term can be interpreted as inserting a mesh of time-oriented Wilson lines $\mathcal{W}=\exp(i\int a_t dt)$ in the underlying $\mathbb{Z}_2$ gauge theory. The equation of motion of $b_\mu$ is not modified by the insertions so $\mathcal{W}_C=\exp(i\oint_C a)$ remains topological and generates a topological $\mathbb{Z}_2^{(a)}$ 1-form symmetry. On the other hand, while the equations of motion of $a_x$ and $a_y$ remain intact, the equation of motion of $a_t$ is modified to
\beq
\partial_x b_y-\partial_y b_x=\frac{1}{\ell^2}.
\eeq
It implies that the $\mathbb{Z}_2^{(b)}$ 1-form symmetry operator $\mathcal{T}_C=\exp(i\oint_C b)$, when defined on a spatial loop $C$, is not topological, but instead depends on the area $A$ enclosed by $C$ as
\begin{equation}
    \mathcal{W}(C)=e^{{i\pi A}/{\ell^2}}.
\end{equation}
This reproduces the behavior of the $\mathbb{Z}_2^{(b)}$ 1-form symmetry operators on the lattice and just as on the lattice, we can redefine the symmetry operators in the  effective field theory by a $c$-number as
\beq
\widetilde{\mathcal{W}}(C)=\exp\left[i\oint_C \left(b-\frac{1}{\ell^2}xdy\right)\right],
\eeq
so that the $\mathbb{Z}_2^{(b)}$ 1-form symmetry becomes topological. The topological $\mathbb{Z}_2^{(a)}\times\mathbb{Z}_2^{(b)}$ 1-form symmetry in the low-energy effective field theory implies the robustness of the four-fold ground state degeneracy.

The effective field theory in Eq.~\eqref{eq:Z2_gauge_theory_mesh} is in fact related to the original $\mathbb{Z}_2$ gauge theory by the redefinition 
$b_y\rightarrow b_y+x/\ell^2$,
which is the field theory analog of the unitary transformation that relates the odd toric code to the original toric code on the lattice.

\subsection{Toric code without plaquette terms}
\label{sec:toric_code_no_plaquette}

Next, we consider the second modification of the toric code code Hamiltonian in Eq.~\eqref{eq:toric_code}, where we fine tune $J_2=0$ while keeping $J_1>0$.
This Hamiltonian still has the $\mathbb{Z}_2^{(a)}\times\mathbb{Z}_2^{(b)}$ lattice 1-form symmetry generated by $A_v$ and $B_p$, respectively. In addition to this 1-form symmetry, there is an extensive symmetry generated by $Z_i$ on every links. We will see that, in this fine-tuned model, unlike the toric code, the lattice $\mathbb{Z}_2^{(b)}$ 1-form symmetry remains non-topological in the low-energy effective field theory.

This fine-tuned model has a drastically different spectrum compared to the toric code. The spectrum remains gapped but has an extensive ground state degeneracy of $2^{L_xL_y+1}$. These ground states satisfy $A_v=1$ but can have distinct eigenvalues under $B_p$. Thus, in the ground state subspace, the $\mathbb{Z}_2^{(b)}$ lattice 1-form symmetry remains non-topological, while the $\mathbb{Z}_2^{(a)}$ lattice 1-form symmetry  becomes topological. This is consistent with the fact that the four-fold ground state degeneracy imposed by the non-commutative algebra in Eq.~\eqref{eq: toric code commutator} is not robust. We can lift the ground state degeneracy completely by adding a weak local perturbation $\epsilon\sum_i X_i$ that breaks the $\mathbb{Z}_2^{(b)}$ 1-form symmetry.

At low energies, the lattice model is described by the following effective field theory Lagrangian
\begin{equation}
    \begin{aligned}
    &\mathcal{L}=\mathcal{L}_{\mathbb{Z}_2}+\mathcal{L}_{\text{M}},
    \\
    &\mathcal{L}_{\text{M}}=\frac{2}{2\pi} B (\partial_t \phi-a_t),
    \end{aligned}
\end{equation}
which can be interpreted as coupling the low-energy effective field theory of the toric code, i.e.~$\mathbb{Z}_2$ gauge theory $\mc{L}_{\mathbb{Z}_2}$ in Eq.~\eqref{eq:Z2_gauge_theory}, to a special matter system with local conservation laws described by $\mathcal{L}_{\text{M}}$. This is similar to the low-energy effective field theory of the $\R$ Kitaev model discussed in Eq.~\eqref{eq:CS+M}. Here, $\phi$ is a compact scalar with position-dependent $2\pi$-periodicity, $\phi\sim\phi +2\pi [\theta(x-x_1)-\theta(x-x_2)][\theta(y-y_1)-\theta(y-y_2) ]$ where $\theta(y)$ is the Heaviside theta function, and $x_{1,2}$ and $y_{1,2}$ can take arbitrary values. $B$ is another compact scalar with position-dependent periodicity, $B\sim B+2\pi\delta(x-x_0)\delta(y-y_0)$, where $x_0$ and $y_0$ can take arbitrary values.  
The matter system when not coupled to the gauge field $a_t$ describes a collections of decoupled qubits and each of them has two independent $\mathbb{Z}_2$-valued locally conserved quantity, $e^{i\phi}$ and $e^{i\int_M B}$, where $M$ is an open region.\footnote{Following a similar analysis as in footnote 7, summing over the windings of $B$ and $\phi$ in the time direction constrains $e^{i\phi}=\pm1$ and $e^{i\int_M B}=\pm1$, respectively.} The coupling has the effect of relaxing the Gauss law imposed by the equation of motion of $a_t$, thereby mimicking the presence of ground states with different $B_p$ eigenvalues in the lattice model.

The full theoru has the following gauge symmetry
\begin{equation}
    a_\mu\rightarrow a_\mu+\partial_\mu\alpha, \quad b_\mu\rightarrow b_\mu+\partial_\mu\beta,\quad\phi\rightarrow \phi+\alpha.
\end{equation}
Using this gauge symmetry, we can gauge fix $\phi=0$ and then integrate out $B$ setting $a_t=0$. In the end, the full Lagrangian simplifies to
\begin{equation}
    \mathcal{L}=\frac{2}{2\pi}\big[a_y\partial_t b_x+b_y\partial_t a_x +b_t(\partial_xa_y-\partial_ya_x)\big]\label{eq:EFT_toric_code_fine_tune}
\end{equation}
where $b_\mu$ remains to be a $U(1)$ gauge field while $a_\mu$ becomes a vector field without any gauge symmetry. The equations of motion of $b_\mu$,
\beq
\partial_ta_x=\partial_t a_y=\partial_xa_y-\partial_ya_x=0,
\eeq
imply that the spatial Wilson loops $\mathcal{W}_C=\exp(i\oint_C a)$ supported on a spatial loop $C$ is conserved and topological. Thus, these loop operators generate a topological $\mathbb{Z}_2^{(a)}$ 1-form symmetry in the low-energy effective field theory, agreeing with the expectation from the lattice model. In addition to these loop operators, these equations of motion also implies the conservation of open line operators $\exp(i\int_\gamma a)$ with $\gamma$ an open path.  Note that $a$ is a vector field without any gauge symmetry, so these open line operators are gauge invariant. These open line operators are the low-energy images of the additional conserved quantity on every links generated by $Z_i$ in the lattice model. 
On the other hand, the equations of motion of $a_\mu$,
\beq
\partial_t b_x-\partial_x b_t=\partial_t b_y-\partial_y b_t=0.
\eeq
imply that the spatial Wilson loops $\mathcal{T}_C=\exp(i\oint_C b)$ supported on a spatial loop $C$ is conserved. However, unlike in the pure $\mathbb{Z}_2$ gauge theory in Eq.~\eqref{eq:Z2_gauge_theory}, the condition $\partial_x b_y-\partial_y b_x=0$ is missing, so these loop operators are not topological and hence generate a non-topological $\mathbb{Z}_2^{(b)}$ 1-form symmetry in the low-energy effective field theory, agreeing with the expectation from the lattice model. The local operator $\mathcal{O}=a_x^2+a_y^2$ is charged under this non-topological $\mathbb{Z}_2^{(b)}$ 1-form symmetry and adding it to the Lagrangian completely gaps out the theory to a unique gapped ground state.

\subsection{Bravyi-Hastings-Michalakis's non-robust toric code} \label{app: non-robust toric code}

We now turn to discuss another fine-tuned modification of toric code proposed in Sec.~2.4 of Ref.~\cite{Bravyi2010} by  Bravyi, Hastings and Michalakis. Similar to the toric code, this is a gapped lattice model with a $\mathbb{Z}_2^{(a)}\times\mathbb{Z}_2^{(b)}$ lattice 1-form symmetry, and its ground states are identical to those of the toric code. However, interestingly, the ground states are not robust against weak local perturbations. We will demonstrate this non-robustness is related to the existence of a symmetry sector  $\mathfrak{s}$ where $\Delta(\mathfrak{s})/N(\mathfrak{s})\rightarrow0$ in the thermodynamic limit. Here, $\Delta(\mathfrak{s})$ denotes the lowest excitation energy of this sector compared to the global ground states and $N(\mathfrak{s})$ is the number of contractible loop eigenvalues that differ from the ground-state sector. We will further discuss the low-energy effective field theory for this model and show that it has a non-topological 1-form symmetry. This specific example motivates our formulation of a general criterion for a topological higher-form symmetry to arise at low energies, as detailed in Sec.~\ref{sec: compactness}.

The model is defined on a square lattice, where a qubit is placed at each link of the lattice, just like the standard toric code model in Ref.~\cite{Kitaev2003}. However, its Hamiltonian is different from the standard toric code model, and it is
\beq \label{eq: non-robust toric code}
H=-J_1\sum_vA_v-J_2B_{p_0}-J_3\sum_{\langle p_1,p_2\rangle}B_{p_1}B_{p_2}
\eeq

The coupling constants $J_{1,2,3}>0$. The third term is a ferromagnetic Ising coupling between any pair of adjacent plaquettes, while $p_0$ labels a fixed plaquette and is not summed over. Same as the toric code, this model has a $\mathbb{Z}_2^{(a)}\times \mathbb{Z}_2^{(b)}$ lattice 1-form symmetry generated by the vertex $A_v$ and plaquette $B_p$ operators, respectively. 

For simplicity, suppose that $J_1>J_3>J_2$ and the system is on a torus with lengths $L_1$ and $L_2$ along the two directions, where $L_1L_2$ is even. Then the Hamiltonian has 4 ground states, which satisfy
\beq \label{eq: original ground states}
A_v=1,\quad B_p=1
\eeq
for all vertices $v$ and all plaquettes $p$. These ground states are identical to those of the standard toric code \cite{Kitaev2003}. Moreover, it is straightforward to check that this Hamiltonian is gapped with a gap
\beq
\Delta=2J_2,
\eeq
and the lowest-energy excited states are also four-fold degenerate and satisfy
\beq \label{eq: new ground states}
A_v=1,\quad B_p=-1
\eeq
for all vertices $v$ and all plaquettes $p$. For the symmetry sector $\mathfrak{s}$ of these lowest-energy excited states,  $\Delta(\mathfrak{s})=2J_2$ and $N(\mathfrak{s})=L_1L_2$ (recall $N(\mathfrak{s})$ is defined as the number of contractible loop eigenvalues that differ from those of the ground state). Therefore, 
$\Delta(\mathfrak{s})/N(\mathfrak{s})=2J_2/(L_1L_2)\rightarrow 0$ in the thermodynamic limit. Furtheremore, since $N(\mathfrak{s})=L_1L_2$, this symmetry sector lies in a superselection sector distinct from the ground state, \ie no local operators with finite support can connect the two.

Now consider the following local perturbation, 
\beq \label{eq: perturbation non-robust toric code}
V=J_4\sum_pB_p.
\eeq
As long as $J_4>J_2/(L_1L_2)$, which only requires an infinitesimal strength in the thermodynamic limit, the states satisfying Eq.~\eqref{eq: new ground states} become the new ground states, which are identical to those of the odd toric code. This sudden change of the ground states indicates the non-robustness of the lattice model and is related to the violation of the condition in  Eq.~\eqref{eq: topological condition}, which we will elaborate more in the next section.

Below we discuss the low-energy effective field theory of the lattice model in Eq.~\eqref{eq: non-robust toric code}. Because the model is gapped and its ground states are identical to those of the standard toric code, which is described by a $\z_2$ gauge theory, it might be tempting to conclude that the low-energy effective field theory of this model is also a $\z_2$ gauge theory in \eqref{eq:Z2_gauge_theory}. However, this proposal is incorrect, because this presumed effective field theory does not capture the non-robustness of the lattice model. A lattice-scale perturbation Eq.~\eqref{eq: perturbation non-robust toric code} would induce a local perturbation in the low-energy effective field theory but there are no local operators in the $\z_2$ gauge theory that can trigger a sudden change of the ground states discussed above. The proper low-energy effective field theory should actually capture both the states satisfying Eq.~\eqref{eq: original ground states} and states satisfying Eq.~\eqref{eq: new ground states}. 

To construct the low-energy effective field theory, we start with the case where $J_2=0$. In this particular case, there is an accidental $\mathbb{Z}_2$ symmetry that maps $B_p\rightarrow -B_p$. In this case, the effective Lagrangian is 

\begin{equation}
\mc{L}=\mc{L}_{\mathbb{Z}_2}+\mathcal{L}_{\text{SSB}}-\frac{2}{2\pi\ell^2} a_t\varphi,\label{eq:EFT_BHM}
\end{equation}
where $\ell$ is a length scale such that $\ell^2$ is the size of the unit cell, $\mc{L}_{\mathbb{Z}_2}$ is the $\mathbb{Z}_2$ gauge theory Lagrangian in Eq.~\eqref{eq:Z2_gauge_theory} and $\mathcal{L}_{\text{SSB}}$ is a $\mathbb{Z}_2$ symmetry breaking Lagrangian given by
\begin{equation}
    \mathcal{L}_{\text{SSB}}=\frac{2}{4\pi}\epsilon^{\mu\nu\lambda} B_{\mu\nu}\partial_\lambda \varphi.
\end{equation}
Here, $B_{\mu\nu}$ is a $U(1)$ 2-form gauge field and $\varphi$ is a $U(1)$ compact scalar with periodicity $\varphi\sim\varphi+2\pi$. They are coupled through another BF-type Lagrangian, which describe a symmetry breaking phase for the accidental $\z_2$ symmetry that maps $B_p\rightarrow -B_p$. The equation of motion of $B_{\mu\nu}$, $\partial_\lambda\varphi=0$, constrains $\varphi$ to be a constant. Furthermore, summing over flux sectors of $B_{\mu\nu}$ with $\oint dB\in 2\pi\mathbb{Z}$ further constrains $e^{i\varphi}=\pm1$, 

so the BF theory has two vacua from the $\mathbb{Z}_2$ symmetry breaking, which are distinguished by their expectation value of the order parameter $\langle e^{i\varphi}\rangle =\pm 1$.

The full Lagrangian in Eq.~\eqref{eq:EFT_BHM} couples the $\mathbb{Z}_2$ gauge theory and the $\mathbb{Z}_2$ symmetry breaking theory. This Lagrangian has the following gauge symmetry:
\begin{equation}
\begin{aligned}
    &a_\mu\rightarrow a_\mu+\partial_\mu\alpha,
    \\
    &b_\mu\rightarrow b_\mu+\partial_\mu\beta,
    \\
    &B_{\mu\nu}\rightarrow B_{\mu\nu}+\partial_\mu\xi_\nu-\partial_\nu\xi_\mu-\frac{\alpha}{\ell^2}\epsilon_{\mu\nu t},
\end{aligned}
\end{equation}
where $\alpha$ and $\beta$ are $U(1)$ compact scalar and $\xi_\mu$ is $U(1)$ 1-form gauge field. With this coupling, the equation of motion of $a_t$ is modified to
\begin{equation}
    \partial_x b_y-\partial_y b_x=\frac{1}{\ell^2}\varphi.
\end{equation}
It implies that the $\mathbb{Z}_2^{(b)}$ 1-form symmetry operator $\mathcal{T}_C=\exp(i\oint_C b)$ defined on spatial loops $C$ is not topological but instead depends on the shape of $C$:
\begin{equation}
    \mathcal{T}_C=\exp\left(\frac{i}{\ell^2}\int_\Omega \varphi\right),
\end{equation}
where $\Omega$ is the region enclosed by the closed loop. In the vacuum $e^{i\varphi}=1$, $\mathcal{T}_C=1$ and in the other vacuum $e^{i\varphi}=-1$, $\mathcal{T}_C=\exp( i  \pi A/\ell^2)$ with $A$ the area enclosed by $\gamma$. If $\ell^2$ is chosen to be the area of the unit cell on the lattice, then $A/\ell^2$ can only take integer values and thus $\mathcal{T}_C=\pm 1$ depending on whether the number of unit cells $C$ enclosed is even or odd. This reproduces the property of the lattice mode.

Interestingly, although $\mathbb{Z}_2^{(b)}$ 1-form symmetry operator $\mathcal{T}_C$ is not topological, we can define a topological and conserved loop operator in this low-energy effective field theory
\begin{equation}
\widetilde{\mathcal{T}}_C=\exp\left[i\oint_C \left(b-\frac{1}{\ell^2} \varphi \,xdy\right)\right],    
\end{equation}
which generates topological $\mathbb{Z}_2$ 1-form symmetry in the effective field theory. On a torus, the non-contractible symmetry operators of this topological $\mathbb{Z}_2$ 1-form symmetry and the $\mathbb{Z}_2^{(a)}$ 1-form symmetry do not commute
\begin{equation}
\widetilde{\mathcal{T}}_x \mathcal{W}_y  =-\mathcal{W}_y \widetilde{\mathcal{T}}_x,\quad  \mathcal{W}_x \widetilde{\mathcal{T}}_y   =-\widetilde{\mathcal{T}}_y \mathcal{W}_x,
\end{equation}
which indicates a mixed-anomaly between the two topological 1-form symmetries and imposes a robust four-fold degeneracy.
This topological $\mathbb{Z}_2$ 1-form symmetry, however, has not lattice counterpart since the redefinition is not by a $c$-number but instead involves the field $\varphi$.

When $J_2\neq 0$, the following localized term is added to the effective Lagrangian 
\beq
\delta\mc{L}=\lambda_2\cos\varphi(\vec x)\,\delta^{(2)}(\vec x-\vec x_0)
\eeq
where $\vec x_0$ is a fixed position. This term favors $e^{i\varphi}=1$ ($e^{i\varphi}=-1$) if $\lambda_2>0$ ($\lambda_2<0$), which corresponds to $J_2>0\;(J_2<0)$. With $\delta\mc{L}$ included, states with both $\partial_xb_y-\partial_yb_x=0$ and $\partial_xb_y-\partial_yb_x=1/\ell^2$ remain in the effective field theory, so the 1-form symmetry is non-topological.

The perturbation Eq.~\eqref{eq: perturbation non-robust toric code} corresponds to a term
\beq
\delta\mc{L}'=\lambda_4\cos\varphi.
\eeq
With this term included, for $\lambda_{4}>0$ ($\lambda_{4}<0$), the states with $e^{i\varphi}=1$ is favored (removed) and the states with $e^{i\varphi}=-1$ are removed (favored) in the low-energy theory, which corresponds to $J_4 < 0 $ ($J_4>0$). This reproduces the drastic change of the ground states in the lattice model and shows that the model is fine-tuned.

\section{Compact and non-compact higher-form symmetries} \label{sec: compactness}

{With the concrete models studied in Sec.~\ref{sec: harmonic oscillator model} and Sec.~\ref{sec: lattice model with compact Z2 1-form}, we are motivated to address a general question:~When does a higher-form symmetry in a lattice system flow to a topological higher-form symmetry at low energies?} In this section, we argue that the answer lies in the compactness of the symmetry. We first establish a necessary condition for a topological higher-form symmetry to arise at low energies and then argue that, \emph{generically} a non-compact higher-form symmetry in a lattice system does not flow to a topological higher-form symmetry. On the other hand, compact higher-form symmetries are \emph{generally} consistent with this necessary condition, strongly suggesting that compact higher-form symmetries do flow to topological higher-form symmetries. Here, compact ones include $\z_n$ and $\U$ higher-form symmetries, and non-compact ones include the $\R$ 1-form symmetry discussed in Sec.~\ref{sec: harmonic oscillator model} and the $\z$ 1-form symmetry discussed in Appendix \ref{app: Z 1-form}.

\subsection{Energy spectrum}

\begin{figure}[h!]
    \centering
    \includegraphics[width=\linewidth]{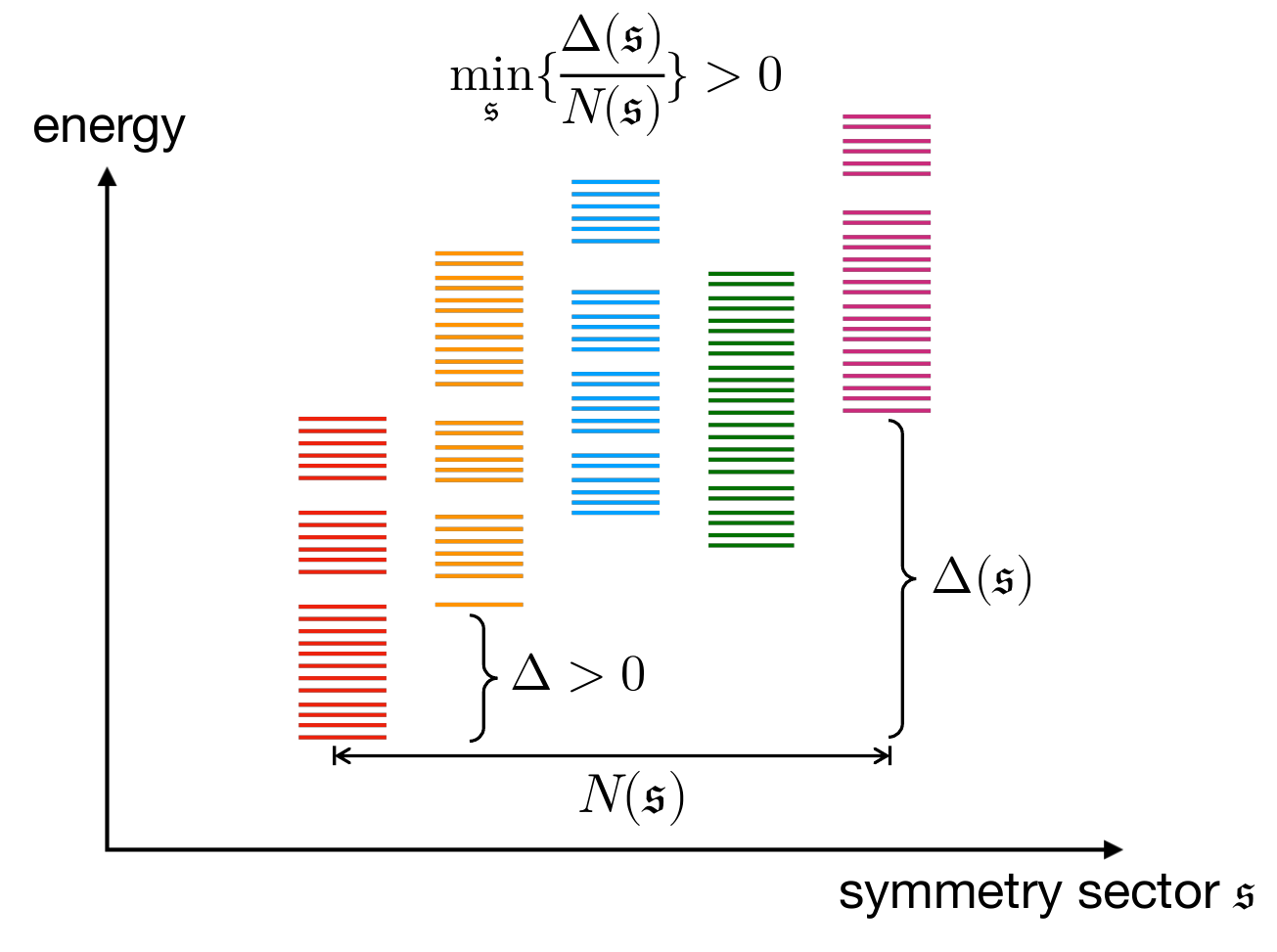}
    \caption{Schematic energy spectrum of a lattice system with a higher-form symmetry that flows to a topological higher-form symmetry at low energies, and satisfies Eq.~\eqref{eq: topological condition}. The horizontal axis represents symmetry sectors characterized by the collection of eigenvalues under all symmetry operators supported on contractible loops. Different sectors are labeled by different colors. Each short line represents an energy eigenstate. $\Delta(\mathfrak{s})$ is the energy difference between the lowest-energy state in sector $\mathfrak{s}$ and the true ground state, $\Delta \equiv \min_\mathfrak{s}\{\Delta(\mathfrak{s})\}$, and $N(\mathfrak{s})$ is the number of contractible loop eigenvalues differing from those in the ground-state sector. The spectrum in the ground-state symmetry sector (labeled red) can be either gapped or gapless.
    } 
    \label{fig:energy_gap_spectrum}
\end{figure}

\begin{figure}[h!]
    \centering
    \includegraphics[width=\linewidth]{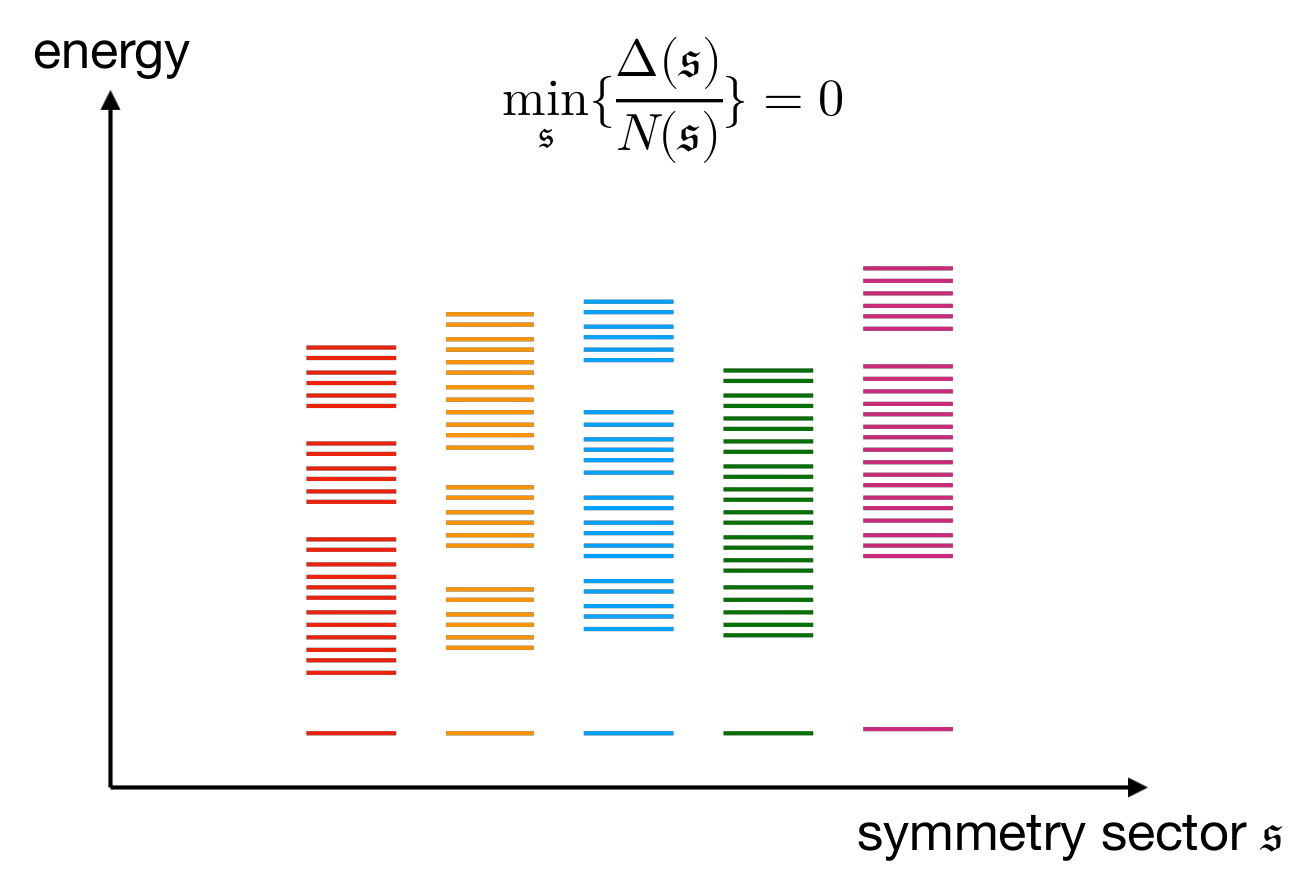}
    \caption{Schematic energy spectrum of a lattice system with a higher-form symmetry that does not flow to a topological higher-form symmetry, and violates Eq.~\eqref{eq: topological condition}. In this figure, the energy gap within each symmetry sector may be zero or non-zero, and it is not meant to be reflected in the figure.} 
    \label{fig: possibly spurious spectrum}
\end{figure}

Our argument is based on the examination of the energy spectrum of a lattice Hamiltonian with a 1-form symmetry. Because the 1-form symmetry operators supported on contractible loops all commute, the energy eigenstates can be chosen to be simultaneous eigenstates of all these symmetry operators. In the example of model in Eq.~\eqref{eq:free_boson_Hamiltonian}, these symmetry operators are precisely the $Q_l$'s in Eq.~\eqref{eq: generators}. Each group of energy eigenstates that share the same eigenvalues under all these symmetry operators form a symmetry sector. In general, the energy spectrum of the Hamiltonian may take the form of either Fig.~\ref{fig:energy_gap_spectrum} or Fig.~\ref{fig: possibly spurious spectrum}. These two types of spectrum are distinguished by the following condition in the thermodynamic limit:
\begin{equation}\label{eq: topological condition}
    \lim_{L\rightarrow\infty}\min_\mathfrak{s}\left\{\frac{\Delta(\mathfrak{s})}{N(\mathfrak{s})}\right\} > 0, 
\end{equation}
with Fig.~\ref{fig:energy_gap_spectrum} satisfying this condition and Fig.~\ref{fig: possibly spurious spectrum} violating this condition. Here $\lim_{L\rightarrow\infty}$ stands for the thermodynamic limit, and we have introduced two quantities $\Delta(\mathfrak{s})$ and $N(\mathfrak{s})$ that characterize every symmetry sector $\mathfrak{s}$: $\Delta(\mathfrak{s})$ is the energy difference between the lowest-energy state of $\mathfrak{s}$ and the overall ground states, and $N(\mathfrak{s})$ is the number of elementary contractible loop operators that have eigenvalues different from those of the ground-state symmetry sector. In Eq.~\eqref{eq: topological condition}, the minimization is over all possible symmetry sectors except for the ground-state sector. Note that $N(\mathfrak{s})\geqslant 1$ by definition. In the example of the $\R$ Kitaev model in Sec.~\ref{sec: harmonic oscillator model} and the fine-tuned model in Sec.~\ref{sec:toric_code_no_plaquette}, $\Delta\equiv\min_\mathfrak{s}\{\Delta(\mathfrak{s})\}=0$ for any system size, hence Eq.~\eqref{eq: topological condition} is violated and the spectrum corresponds to Fig.~\ref{fig: possibly spurious spectrum}. In contrast, for the famous toric code model \cite{Kitaev2003}, $\Delta(\mathfrak{s})\sim N(\mathfrak{s})$ (as the excitation energy is proportional to the number of flipped plaquettes and vertices by a size-independent constant), so Eq.~\eqref{eq: topological condition} is satisfied.

Next, we argue that Eq.~\eqref{eq: topological condition} is a necessary condition for the lattice higher-form symmetry to flow to a topological higher-form symmetry in the low-energy effective field theory.  
We first note that Eq.~\eqref{eq: topological condition} implies $\Delta\equiv\min_\mathfrak{s}\{\Delta(\mathfrak{s})\}>0$, and let us begin by explaining why this is needed. Consider a state $|\psi\rangle$ from the same superselection sector as a vacuum state in a field theory with a topological higher-form symmetry. 

For any symmetry operator $W_l$ supported on a contractible loop, ${\langle\psi|W_l|\psi\rangle}{}=\langle\psi|\psi\rangle$ because $W_l$ can be shrunk and disappear. Suppose $O$ is a local operator. Then ${\langle\psi|O^\dag W_lO|\psi\rangle}=\langle\psi|O^\dag O|\psi\rangle$, where we have used that $W_l$ can be deformed, shrunk and disappear. Therefore, symmetry operators supported on contractible loops just act as identity on all the states from the same superselection sector as a vacuum state. 

Coming back to the spectra sketched in Fig.~\ref{fig:energy_gap_spectrum} and Fig.~\ref{fig: possibly spurious spectrum}, only the part in Fig.~\ref{fig:energy_gap_spectrum} with energy scales below $\Delta > 0$ satisfies this condition.{\footnote{The equation ${\langle\psi|W_l|\psi\rangle}{}=\la\psi|\psi\ra$ appears to further require that the low-energy states must have eigenvalue 1 under all $W_l$ operators. However, in Fig.~\ref{fig:energy_gap_spectrum}, the energy eigenstates with energies smaller than $\Delta$ may have $W_l$ eigenvalues different from 1. But this is not an issue, because in this case, one can always redefine the symmetry operators by a phase factor, so that the eigenvalues under the new symmetry operators are 1. Then one can regard these new symmetry operators as the ones that can be deformed and shrunk in the low-energy effective field theory. This situation is analogous to that of the odd toric code discussed in Sec.~\ref{sec:odd_toric_code}.}}  Therefore, if the spectrum behaves as in Fig.~\ref{fig: possibly spurious spectrum}, the lattice higher-form symmetry cannot flow to a topological higher-form symmetry at low energies.

However, just having a spectrum of the form of Fig.~\ref{fig:energy_gap_spectrum} with the gap $\Delta>0$ is still not enough for the lattice higher-form symmetry to flow to a topological higher-form symmetry in the low-energy effective field theory, and Eq.~\eqref{eq: topological condition} is further required. This is illustrated by the non-robust toric code discussed in Sec.~\ref{app: non-robust toric code}, which has a spectrum as in Fig.~\ref{fig:energy_gap_spectrum} and $\min_\mathfrak{s}\{\Delta(\mathfrak{s})/N(\mathfrak{s})\}=0$ in the thermodynamic limit.

To understand why Eq.~\eqref{eq: topological condition} is necessary for a topological higher-form symmetry at low energies, let us consider violating this condition by having an excited-state sector $\mathfrak{s^*}$  with $\Delta(\mathfrak{s^*})>0$ but $\Delta(\mathfrak{s^*})/N(\mathfrak{s^*})$=0 in the thermodynamic limit. This suggests that, while the energy difference compared to the ground state is non-zero, it is unnaturally small such that by adding an arbitrarily weak local perturbation (say, as a sum of contractible local loop operators, see Eqs.~\eqref{eq: perturbation discrete} and ~\eqref{eq: perturbation continuous} in the later discussion) it is possible to turn this excited-state sector into the ground-state sector, and turn the original ground-state sector into an excited-state sector. As such, both sectors should be included in the low-energy effective field theory. Now because the low-energy effective field theory includes states that have different eigenvalues for the contractible loop operators, its higher-form symmetry is not topological. This phenomenon is demonstrated in the example in Sec.~\ref{app: non-robust toric code}.

In the rest of this section, we will show that the energy spectrum of a lattice system with a non-compact (compact) 1-form symmetry 
generically takes the form of Fig.~\ref{fig: possibly spurious spectrum} (Fig.~\ref{fig:energy_gap_spectrum}), with Eq.~\eqref{eq: topological condition} violated (satisfied) 
in the thermodynamic limit. In fact, for a generic Hamiltonian with a non-compact 1-form symmetry, we will argue that $\Delta=0$ for any system size, which implies the violation of Eq.~\eqref{eq: topological condition}. These results imply that, generically, a higher-form symmetry in a lattice system flows (does not flow) to a topological higher-form symmetry if it is compact (non-compact).

\subsection{Non-compact 1-form symmetries} \label{subsec: noncompact}

In this subsection, we argue that the spectrum of a lattice Hamiltonian with a non-compact 1-form symmetry generically takes the form of Fig.~\ref{fig: possibly spurious spectrum}, particularly with a vanishing gap $\Delta=0$. To understand it, the key is that the eigenvalues of the contractible loop operators for such a symmetry take continuous values.

To be concrete, consider a bosonic model based on the setup in Sec.~\ref{sec: harmonic oscillator model}, which could be interacting, for example, by appropriately choosing the functions $f_{X, Y, Z}$. As discussed at the beginning of Sec.~\ref{subsec: free boson spectrum}, the $\R$ lattice 1-form symmetry implies the presence of $L_1L_2-1$ commuting conserved quantities, associated with contractible loop operators. 
Denoting the eigenvalues of these contractible loop operators $Q_l$ by $q_l$, the Hilbert space $\mc{H}$ and the Hamiltonian $H$ can be decomposed into symmetry sectors where $\{q_{l}\}$ take different values, \ie

\beq
\begin{split}
\mc{H}=\bigoplus_{\{q_{l}\}}\mc{H}_{\{q_{l}\}},\;\;\;H=\sum_{\{q_{l}\}}H_{\{q_{l}\}}.
\end{split}
\eeq
Importantly, the values of $\{q_{l}\}$ in each symmetry sector serve as parameters of the Hamiltonian in this symmetry sector, $H_{\{q_{l}\}}$.

For a non-compact higher-form symmetry, the conserved quantities ${q_l}$ take continuous values. For example, in the case of the $\mathbb{R}$ 1-form symmetry discussed in Sec.~\ref{sec: harmonic oscillator model}, the eigenvalues of $Q_l$ can be arbitrary real numbers. Since the lowest energy $E({q_l})$ in each symmetry sector is generically a smooth function of ${q_l}$, the resulting spectrum typically exhibits the behavior shown in Fig.~\ref{fig: possibly spurious spectrum}, unless the Hamiltonian is fine-tuned so that $E({q_l})$ becomes discontinuous in ${q_l}$ with a nonzero gap $\Delta>0$.

This implies that an $\mathbb{R}$ lattice 1-form symmetry generically flows to a non-topological $\mathbb{R}$ 1-form symmetry at low energies and, consequently, is not robust against weak local perturbations. This conclusion is consistent with Ref.~\cite{Hastings2005}, which argued that an $\mathbb{R}$ lattice 1-form symmetry in a lattice model that naively flows to a Maxwell theory with gauge group $\mathbb{R}$ is not robust due to the presence of gapless electric charges. The same reasoning applies to other non-compact higher-form symmetries, such as the $\mathbb{Z}$ 1-form symmetry realized in the lattice Hamiltonians discussed in Appendix~\ref{app: Z 1-form}.

Our result suggests that one cannot physically engineer an $\mathbb{R}$ Chern–Simons theory by just imposing an exact anomalous $\mathbb{R}$ 1-form symmetry in a lattice system. In such systems, the physical Hilbert space admits a tensor-product structure and the Hamiltonian does not have any singular dependence on the eigenvalues of the contractible loop operators. 
Theoretically, one can of course consider a Hamiltonian that contains a Dirac delta function or impose a hard constraint, such as Gauss law constraints in gauge theories, that forces the system to be in a single eigenvalue sector of the contractible loop operators (see, \eg various lattice regularizations of non-compact Chern-Simons theory explored recently in Refs.~\cite{chen2021abelian,jacobson2024canonical,Magdalena2024}), but such systems do not appear to be physically realizable.

Note that our statement here does not immediately forbid the emergence of an $\R$ Chern-Simons theory in a condensed matter system. Instead, our claim is that if an $\R$ Chern-Simons theory emerges in such cases, the $\mathbb{R}$ 1-form symmetry associated with it should not originate from an exact $\mathbb{R}$ lattice 1-form symmetry of the lattice system. An example of emergent $\mathbb{R}$ 1-form symmetries from a discrete system can be found in gapless infinite-component Chern-Simons-Maxwell theories constructed in Ref.~\cite{chen2022gapless}, which is continuous in all directions except for one discrete direction.

\subsection{Compact 1-form symmetries} \label{subsec: generic compact 1-form}

In Sec.~\ref{subsec: noncompact}, we have seen that a lattice Hamiltonian with a non-compact higher-form symmetry generically has a spectrum like Fig.~\ref{fig: possibly spurious spectrum}, due to the fact that the eigenvalues of the symmetry operators on contractible loops can take continuous values. In this subsection, we show that the spectrum of a lattice Hamiltonian with a compact 1-form symmetry generically takes the form of Fig.~\ref{fig:energy_gap_spectrum} with $\Delta>0$, because of the discrete nature of the eigenvalues of the corresponding symmetry operators on contractible loops. 
Furthermore, Eq.~\eqref{eq: topological condition} is generally satisfied unless fine tuned. Examples of such fine-tuned Hamiltonian are discussed in Sec.~\ref{sec:toric_code_no_plaquette} and Sec.~\ref{app: non-robust toric code}.

For concreteness, consider a lattice Hamiltonian with a $\z_n$ 1-form symmetry. Suppose its energy spectrum takes the form as Fig.~\ref{fig: possibly spurious spectrum}, which has $\Delta=0$ in the thermodynamic limit. Then we can add a weak local perturbation that respects the 1-form symmetry to lift the (approximate) degeneracy between the different low-energy symmetry sectors and select a single sector to have the lowest energy with $\Delta>0$, thereby making the energy spectrum behaves as Fig.~\ref{fig:energy_gap_spectrum}. To do so, denote the eigenvalues of this singled out sector under the contractible loop operators $W_l$ by $\lambda_l$, then the perturbation can be taken as
\beq \label{eq: perturbation discrete}
\delta H=-t\sum_{l}(\lambda_{l}^{*}W_{l}+\lambda_{l}W_{l}^{\dagger})
\eeq
with $t>0$. Importantly, for a $\z_n$ 1-form symmetry, $W_l^n=1$, so the eigenvalues of $W_l$ are $n$-th root of unity, which take discrete values. This allows the perturbation to lift the ground state degeneracy. Due to the constraint $\prod_l W_l=1$ on a torus, the first excited sector of the perturbation $\delta H$ has $W_l=\lambda_l$ for most elementary loops, except two of them. On one of these two elementary loops, $W_l=\lambda_l e^{2\pi i/n}$, while on the other $W_l=\lambda_l e^{-2\pi i/n}$. As a result, the perturbation opens a gap at least $\Delta=4t\left(1-\cos\left(\frac{2\pi}{n}\right)\right)>0$ between the singled out sector and the other sectors.

Similarly, if the 1-form symmetry under consideration is a $U(1)$ symmetry, then different sectors can be labelled by the eigenvalues under the $Q_l$ operators, which are the generators of $U(1)$ 1-form symmetry supported on contractible loops. Suppose we want to single out a sector with eigenvalue $n_l$ under $Q_l$ to make it have the lowest energy, the local $U(1)$ 1-form symmetric perturbation can be taken as
\beq \label{eq: perturbation continuous}
\delta H=t\sum_{l}(Q_{l}-n_{l})^{2}
\eeq
where $t>0$. Because $n_l$ is an integer, which again takes discrete values, this perturbation will open a gap $\Delta=2t>0$ between this singled out sector and the other sectors.

Moreover, in addition to making $\Delta>0$, the perturbations in Eqs.~\eqref{eq: perturbation discrete} and \eqref{eq: perturbation continuous} would actually satisfy the condition in Eq.~\eqref{eq: topological condition}. To see it, consider a symmetry sector $\mathfrak{s}$ that has different eigenvalues for a number of $N(\mathfrak{s})$ contractible loop operators, and suppose before adding these perturbations the energy difference between its lowest-energy state and the global ground states is $\Delta_0(\mathfrak{s})$. After Eq.~\eqref{eq: perturbation discrete} (resp.~\eqref{eq: perturbation continuous}) is added, the new energy difference between the lowest-energy state in this sector and the true ground states is at least $\Delta(\mathfrak{s})=\Delta_0(\mathfrak{s})+2tN(\mathfrak{s})\left(1-\cos\left(\frac{2\pi}{n}\right)\right)$ (resp.~$\Delta(\mathfrak{s})=\Delta_0(\mathfrak{s})+tN(\mathfrak{s})$). Therefore, in the thermodynamic limit $\Delta(\mathfrak{s})/N(\mathfrak{s})\sim t>0$. Since the above argument works for any excited-state sector, Eq.~\eqref{eq: topological condition} is satisfied.

Therefore, if the spectrum of a lattice Hamiltonian with a compact 1-form symmetry behaves as Fig.~\ref{fig: possibly spurious spectrum}, there always exists a weak local 1-form-symmetric perturbation that converts it into the form of Fig.~\ref{fig:energy_gap_spectrum}.
In this sense, a lattice system with a compact 1-form symmetry generically satisfies Eq.~\eqref{eq: topological condition}. This conclusion is consistent with the fine-tuned models discussed in Sec.~\ref{sec:toric_code_no_plaquette} and Sec.~\ref{app: non-robust toric code}, where the $\mathbb{Z}_2^{(b)}$ 1-form symmetry flows to a non-topological 1-form symmetry in the low-energy effective field theory. Upon adding the perturbation in Eq.~\eqref{eq: perturbation discrete}, the system can become either a toric code or an odd toric code. In both cases, the $\mathbb{Z}_2^{(b)}$ 1-form symmetry becomes topological in the low-energy effective field theory.

\section{Discussion} \label{sec: discussion}

In this work, we have discussed a concrete example of lattice system with an $\R$ lattice 1-form symmetry, and showed that this symmetry does not flow to a topological $\R$ 1-form symmetry in the low-energy effective field theory that we have constructed. On the other hand, via an in-depth analysis of various modifications of the toric code, we have illustrated that an $\mathbb{Z}_2$ lattice 1-form symmetry does become topological in the low-energy effective theory, except for fine-tuned cases that we have clarified. We have further established a necessary condition, see Eq.~\eqref{eq: topological condition}, for a lattice higher-form symmetry to flow to a topological higher-form symmetry at low energies. From this, we have argued that a non-compact lattice higher-form symmetry generically flows to a non-topological higher-form symmetry. Compact lattice higher-form symmetries, on the other hand, are generically consistent with this condition.

We remark that we have not rigorously established that the condition in Eq.~\eqref{eq: topological condition} is sufficient to ensure that the lattice higher-form symmetry flows to a topological higher-form symmetry in the low-energy effective field theory, although we expect this to be true. An important consequence of flowing to a topological higher-form symmetry is the robustness of the higher-form symmetry. Namely, even when it is explicitly broken at the lattice scale by arbitrarily weak local perturbations, the higher-form symmetry re-emerges at low energies. Therefore, to prove the sufficiency of the condition in Eq.~\eqref{eq: topological condition}, one has to at least show that a model obeying Eq.~\eqref{eq: topological condition} still has an approximate higher-form symmetry at low energies under a generic weak local perturbation. As a first step in this direction, one needs to prove that the energy gap $\Delta$ in Fig.~\ref{fig:energy_gap_spectrum} remains finite for any weak local perturbations that respect the original higher-form symmetry in the unperturbed Hamiltonian. However, even this first step is challenging, as the robustness of the energy gap of a generic quantum many-body Hamiltonian is not fully understood, although for special Hamiltonians, such as free-fermion Hamiltonians and some frustration-free Hamiltonians, this robustness can be proved \cite{Bravyi2010, Bravyi2010a}. We leave these important open problems to the future studies.

Moreover, checking the condition in Eq.~\eqref{eq: topological condition} may be challenging for a generic microscopic Hamiltonian with a lattice version of higher-form symmetry, so it is interesting to develop other analogous conditions that are more practical to use.

\begin{acknowledgements}

We thank Han Ma for helpful discussions and for a related collaboration. Research at Perimeter Institute is supported in part by the Government of Canada
through the Department of Innovation, Science and Industry Canada and by the Province of Ontario through
the Ministry of Colleges and Universities.
RL is also supported by the Simons Collaboration on Global Categorical Symmetries through Simons Foundation grant 888996. 
PMT is supported by a postdoctoral research fellowship at the Princeton Center for Theoretical Science and a Croucher Fellowship for Postdoctoral Research.
HTL is supported by the U.S. Department of Energy, Office of Science, Office of High Energy Physics of U.S.~Department of Energy under grant Contract Number DE-SC0012567 (High Energy Theory research), by the Packard Foundation award for Quantum Black Holes from Quantum Computation and Holography and by the Simons Investigator Award No.~926198. 
LZ is supported by the National University of Singapore start-up
grants A-0009991-00-00 and A-0009991-01-00, and by the Ministry of Education, Singapore, under its Academic Research Fund Tier 2 Program (MOE-T2EP50225-0005).
    
\end{acknowledgements}

\appendix

\onecolumngrid

\section{Spectrum of the $\mathbb{R}$ Kitaev model}\label{app:exact_diagonalization}

In this appendix, we present an exact analysis of the energy spectrum of the $\mathbb{R}$ Kitaev model in Eqs.~\eqref{eq:free_boson_Hamiltonian} and \eqref{eq:local_terms_in_Hamiltonian} based on the bosonic Bogoliubov transformation \cite{COLPA1978327, COLPA1986377, COLPA1986417}, which substantiates the $\mathbb{R}$ 1-form symmetry analysis in the main text. We further demonstrate that an explicit breaking of the $\mathbb{R}$ 1-form symmetry indeed lifts the infinite ground state degeneracy on torus. 

Let us express the position and momentum operators for the quantum mechanical particle at each site $\textbf{r}$ in terms of bosonic creation and annihilation operators ($[\alpha_{\textbf{r}},\alpha^\dagger_{\textbf{r}'}]=\delta_{\textbf{r},\textbf{r}'}$):
\begin{equation}
    x_\textbf{r} = \frac{1}{\sqrt{2}}(\alpha_\textbf{r}^\dagger+\alpha_\textbf{r}),\;\; p_\textbf{r}=\frac{i}{\sqrt{2}}(\alpha^\dagger_\textbf{r} - \alpha_\textbf{r}).
\end{equation}
Upon substitution into Eq.~\eqref{eq:free_boson_Hamiltonian} and Fourier transforming to the momentum ($\bk$) space,
\begin{equation}
    \alpha_{\textbf{k},A/B} = \frac{1}{\sqrt{L_1L_2}}\sum_{\textbf{R}_i} \alpha_{\textbf{R}_i, A/B} e^{-i\textbf{k}\cdot \textbf{R}_i},
\end{equation}
with $\textbf{R}_i$ labeling the unit cell and $A/B$ labeling the sublattice, our model is rewritten in the Bogoliubov-de Gennes (BdG) form as $H=\sum_\textbf{k} \ba^\dagger_\bk H_\bk \ba_\bk$, with $\ba^\dagger_\textbf{k}=(\alpha^\dagger_{\textbf{k},A},\alpha^\dagger_{\textbf{k},B},\alpha_{-\textbf{k},B},\alpha_{-\textbf{k},A})$ and
\begin{equation}\label{eq:app HBdG}
    H_\bk=\begin{pmatrix}
    \mathcal{A} & \mathcal{B}\\
    \mathcal{B}^* & \mathcal{A}^*
    \end{pmatrix}
\end{equation}
where 
\begin{subequations}\label{eq:app blockmatrix}
\begin{align}
\mathcal{A}=\frac{1}{2}\begin{pmatrix}
    J_x+J_y(\alpha^2+\beta^2)+J_z & J_x e^{i\bk\cdot\bT_2}+ J_y e^{i\bk \cdot (\bT_2-\bT_1)}(\alpha^2+\beta^2) +J_z \\
    J_x e^{-i\bk\cdot\bT_2}+ J_y e^{i\bk \cdot (\bT_1-\bT_2)}(\alpha^2+\beta^2) +J_z & J_x+J_y(\alpha^2+\beta^2)+J_z
    \end{pmatrix},\\
\mathcal{B}=\frac{1}{2}\begin{pmatrix}
    J_x e^{i\bk\cdot\bT_2}+J_y e^{i\bk\cdot (\bT_2-\bT_1)}(i\alpha+\beta)^2 -J_z & J_x+J_y(i\alpha+\beta)^2-J_z \\ J_x+J_y(i\alpha+\beta)^2-J_z & J_x e^{-i\bk\cdot\bT_2}+J_y e^{i\bk\cdot (\bT_1-\bT_2)}(i\alpha+\beta)^2-J_z
    \end{pmatrix}.
\end{align}
\end{subequations}
Here nonzero constants $\alpha, \beta \in \R$ parametrize the 1-form symmetry, $J_{x,y,z}>0$ are coupling constants and $\{\bT_1,\bT_2\}$ are primitive lattice vectors defined in Fig.~\ref{fig:label_of_links}. 

To obtain the energy spectrum, we consider a transformation to a new set of bosonic operators $\bb_\bk = \mathscr{T}^{-1}\ba_\bk $, where the bosonic commutation relation demands $\mathscr{T}$ to be a \textit{para-unitary} matrix satisfying the condition 
\begin{equation}
    \mathscr{T}^\dagger \hat{\mathscr{I}} \mathscr{T} = \mathscr{T}\hat{\mathscr{I}} \mathscr{T}^\dagger=\hat{\mathscr{I}},
\end{equation}
where $\hat{\mathscr{I}}:=\text{diag}(1,1,-1,-1)$ is the so-called para-unit matrix. Unlike the fermionic case where the spectrum can be directly read off from the unitary diagonalization of $H_\bk$, here for the bosonic case the situation is more subtle as there is no guarantee of a \textit{para-unitary} diagonalization, i.e., $\mathscr{T}^\dagger H_\bk \mathscr{T}$ may never be diagonal. Nonetheless, in a series of works by Colpa \cite{COLPA1978327, COLPA1986377, COLPA1986417}, it was proven that through conjugation by a para-unitary matrix a positive-semidefinite $H_\bk$ can always reach the so-called standardized form:
\begin{equation}
    \mathscr{T}^\dagger H_\bk \mathscr{T}=\mathscr{E}_\bk\equiv \begin{pmatrix}
        E_e &     &    & 0_e &     &   \\
            & I_w &    &     & I'_w &   \\
            &     & 0_z&     &     & 0_z\\
        0_e &     &    & E_e &     &    \\
            & I'_w &    &     & I_w &    \\
            &     & 0_z&     &     & 0_z
    \end{pmatrix},
\end{equation}
with $E_e$ a positive-definite diagonal square matrix of order $e$, $I_w$ the unit matrix of order $w$, $I'_w$ a diagonal matrix of order $w$ with $\pm 1$ diagonal elements, $0_{z}$ a square zero matrix of order $z$, and all other unspecified entries are zero. From $H=\sum_\bk \bb_\bk^\dagger \mathscr{E}_\bk\bb_\bk$, it is straightforward to express $H$ in terms of a set of independent oscillators of the form in Eq.~\eqref{eq:normal_modes}, where the number of type I, II and III modes are $e$, $w$ and $z$, respectively. 

From Eqs.~\eqref{eq:app HBdG} and \eqref{eq:app blockmatrix}, we can explicitly check that $\det H_\bk = 0$ for any $\alpha, \beta \in \R$ and $J_{x,y,z}$. This implies the existence of either type II modes, which give a gapless spectrum, or type III modes, which give an infinite ground state degeneracy. The level-spacing of a type I mode, which is a diagonal element of $E_e$, corresponds to a positive eigenvalue of $\hat{\mathscr{I}}H_\bk$:
\begin{equation}\label{eq: app version type I level spacing}
\begin{split}
    E_1 (\bk) =  2\sqrt{2}\Big( J_x J_y \alpha^2 \cos^2\frac{\bk\cdot\bT_1}{2}+J_y J_z \beta^2 \cos^2 \frac{\bk\cdot\bT_{12}}{2}+J_x J_z \cos^2 \frac{\bk\cdot\bT_2}{2} \Big)^{\frac{1}{2}},
\end{split}
\end{equation}
where $\bT_{12}\equiv \bT_1-\bT_2$. It is obvious that $E_1>0$ for any $\bk$, hence there is $e=1$ number of type I mode for each $\bk$, i.e., in total $L_1L_2$ number of type I modes.\footnote{Recall we have assumed $\alpha,\beta \neq 0$, otherwise there can be $E_1=0$ for specific $\bk$ which leads to additional zero modes and additional conserved quantities. This is considered as fine-tuning and hence ignored in our discussion.} The remaining $L_1L_2$ modes must either be the type II or type III zero modes. In Sec.~\ref{subsec: free boson spectrum}, by the counting of 1-form symmetry generators we have argued that there are altogether $L_1L_2-1$ type II modes and 1 type III mode. Below we check this explicitly. 

According to Theorem 3.11 in Ref.~\cite{COLPA1986377}, the algebraic multiplicity of the zero eigenvalue of $\hat{\mathscr{I}} H_\bk$ and $H_\bk$ are $2(z+w)$ and $2z+w$, respectively. The secular equation for $\hat{\mathscr{I}} H_\bk$ can be analytically solved to give $z+w=1$, while the algebraic multiplicity of the zero eigenvalue of $H_\bk$ (due to its hermiticity) equals the number of associated linearly independent eigenvectors. We find $2z+w= 1$ for $\bk \neq 0$, while $2z+w=2$ for $\bk=0$. Thus, the number of zero modes is given by
\begin{equation}
(z,w)=
    \begin{cases}
    (0,1), \;\;\text{for $\bk \neq 0$,} \\
    (1,0),\;\;\text{for $\bk = 0$},
    \end{cases}
\end{equation}
as claimed in Sec.~\ref{subsec: free boson spectrum}. Furthermore, the two independent zero eigenvectors of $H_{\bk=0}$ can be chosen as $(1,-1,0,0)^T$ and $(0,0,1,-1)^T$, respectively, which implies that the type III zero mode is associated to the oscillator annihilated by $\beta_0=\alpha_{\bk=0,A}-\alpha_{\bk=0,B}$ (it contributes $0\beta^\dagger_0 \beta_0$ to $H$). The corresponding canonical coordinates, $\tilde{x}_0=(\beta_0^\dagger+\beta_0)/\sqrt{2}$ and $\tilde{p}_0=i(\beta^\dagger_0-\beta_0)/\sqrt{2}$, are indeed the two conjugate conserved quantities in Eq.~\eqref{eq:typeIII}.

The $\mathbb{R}$ 1-form symmetry implies infinite degeneracy of all energy levels of the model on a torus. The usual expectation for a lattice version of higher-form symmetry is that it reemerges at low energies, even if it is explicitly broken by weak local perturbations. If this expectation is correct, then the infinite degeneracy on a torus should still be present, at least approximately, when weak local perturbations are introduced to the model in Eqs.~\eqref{eq:free_boson_Hamiltonian} and \eqref{eq:local_terms_in_Hamiltonian}. Below we will show the opposite that this infinite degeneracy is lifted upon adding a weak local perturbation.

We perturb $H$ by adding a harmonic oscillator term to every site: $\epsilon(x^2_\br +p^2_\br-1) = 2\epsilon\alpha^\dagger_\br\alpha_\br$, with $\epsilon>0$. The perturbed BdG Hamiltonian is then $H'_\bk = H_\bk + \epsilon I_4$, and it can be directly seen that $\det H'_\bk \sim \mathcal{O}(\epsilon) > 0$ for all $\bk$, i.e.~all normal modes become type I and the ground state is nondegenerate. More specifically, from the positive eigenvalues of $\hat{\mathscr{I}} H'_\bk$, we find the level-spacing of the two type I modes for each $\bk$: one has the level-spacing in Eq.~\eqref{eq: app version type I level spacing}, to leading order in $\epsilon$, while the other has a level-spacing $E_2 (\bk)$ satisfying
\begin{equation}
\begin{split}
     E^2_1 (\bk) E^2_2 (\bk) = &8\epsilon J_xJ_yJ_z\big(\alpha^2\sin^2{\frac{\bk\cdot\bT_{12}}{2}}+\beta^2 \sin^2{\frac{\bk\cdot\bT_{1}}{2}}\big)+ \\
     &4\epsilon^2 \big[J_xJ_z+J_xJ_y(\alpha^2+\beta^2\sin^2{\frac{\bk\cdot\bT_{1}}{2}})+J_yJ_z(\beta^2+\alpha^2\sin^2{\frac{\bk\cdot\bT_{12}}{2}})\big] + \mathcal{O}(\epsilon^3).
\end{split}
\end{equation}
For a small and finite $\epsilon$, it is clear that $E_2(\bk)$ never vanishes at any $\bk$. $E_2 (\bk)$ is minimized around $\bk=0$, where the level-spacing is at order $\epsilon$. The perturbed free-boson model thus possesses a unique ground state with an energy gap of the order $\epsilon$.

\section{Spectrum of the low-energy effective field theory} \label{app:EFT}

In this appendix, we compute the spectrum of the low-energy effective field theory in Eq.~\eqref{eq:EFT_Lagrangian}, following an analysis similar to that of Appendix~\ref{app:exact_diagonalization}. As we will see, the resulting spectrum exactly matches the low-energy spectrum of the $\R$ Kitaev model.

In the low-energy effective field theory, $A_x$ and $A_y$ are conjugate variables. Upon quantization, we obtain the following commutation relation
\begin{equation}
    [A_x(\mathbf{x}),A_y(\mathbf{x}')]=i\delta^2(\mathbf{x}-\mathbf{x}').
\end{equation}
We can then express these fields in terms of creation and annihilation operators ($[a_{\mathbf{k}},a_{\mathbf{k'}}^\dagger]=\delta^2(\mathbf{k}-\mathbf{k}')$) in momentum $(\mathbf{k})$ space as
\begin{equation}
\begin{aligned}
    &A_x(\mathbf{x})=\frac{1}{\sqrt{2}}\int \frac{d^2\mathbf{k}}{{2\pi}}(a_\mathbf{k}^\dagger e^{-i\mathbf{k}\cdot\mathbf{x}}+a_\mathbf{k} e^{i\mathbf{k}\cdot\mathbf{x}}),
    \\
    &A_y(\mathbf{x})=\frac{i}{\sqrt{2}}\int \frac{d^2\mathbf{k}}{{2\pi}}(a_\mathbf{k}^\dagger e^{-i\mathbf{k}\cdot\mathbf{x}}-a_\mathbf{k} e^{i\mathbf{k}\cdot\mathbf{x}}).
\end{aligned}
\end{equation}
The Hamiltonian
\begin{equation}
H=\int d^2\mathbf{x}\ J(\partial_x A_y-\partial_y A_x)^2,
\end{equation}
when expanded in terms of creation and annihilation operators, takes a BdG form as $ H=\int d^2\mathbf{k}\, \mathbf{a}_\mathbf{k}^\dagger H_\mathbf{k} \mathbf{a}_\mathbf{k}$
with $\mathbf{a}_{\mathbf{k}}^\dagger=(a^\dagger_\mathbf{k},a_{-\mathbf{k}})$ and
\begin{equation}
    H_\mathbf{k}=\frac{J}{2}\left(\begin{array}{cc}
         k_x^2+k_y^2 & -(k_x+ik_y)^2 \\
         -(k_x-ik_y)^2 & k_x^2+k_y^2
    \end{array}\right).
\end{equation}
Since $\det H_\mathbf{k}=0$, the mode at momentum $\mathbf{k}$ must be either type II or type III. To determine the exact type, we consider the eigenvalues of $\hat{\mathscr{I}} H_\mathbf{k}$ and $H_\mathbf{k}$ with $\hat{\mathscr{I}}:=\text{diag}(1,-1)$.  $\hat{\mathscr{I}} H_\mathbf{k}$ has two zero eigenvalues,  while $H_\mathbf{k}$ has a zero eigenvalue and another eigenvalue $2(k_x^2+k_y^2)$, which vanishes only at $\mathbf{k}=0$. Following a similar argument presented in Appendix \ref{app:exact_diagonalization}, we conclude that the mode is type III at $\mathbf{k}= 0$ and type II otherwise. This exactly matches the low-energy spectrum of the $\mathbb{R}$ Kitaev model and therefore substantiates our claim of the low-energy effective theory Lagrangian in Eq.~\eqref{eq:EFT_Lagrangian}.

\section{A Rotor model with a $\z$ 1-form symmetry} \label{app: Z 1-form}

In this appendix, we discuss a lattice rotor model and its $\z$ 1-form symmetry.

This model is also defined on a honeycomb lattice and we put a rotor on each lattice site (see Fig.~\ref{fig:label_of_links}). More precisely, the local Hilbert space at each site is constructed by operators $\{\phi,L\}$ satisfying
\beq\label{eq:compact_CCR}
[\phi,L]=i.
\eeq
Moreover, $\phi$ is compactified in the sense that $\phi\simeq \phi+2\pi$. Given this ambiguity, the operator $e^{in\phi}$ only makes sense when $n\in\z$. Note that Eq.~(\ref{eq:compact_CCR}) indicates that
\beq
\begin{split}
    e^{in\phi}L e^{-in\phi}&=L-n, \\
    e^{in\phi}e^{i\alpha L}e^{-in\phi}e^{-i\alpha L}&=e^{-in\alpha},
\end{split}
\eeq
where $\alpha \in \mathbb{R}$.

Using $\alpha$ and $n$ as two parameters, the $\z$ 1-form symmetry on each elementary loop (\ie hexagon) $l$ can be constructed as follows:
\begin{align}\label{eq:z_1-form_generator}
    W_l(m)=
    \left(e^{-i\alpha L_1}e^{-in\phi_1}e^{i\alpha L_2}e^{in\phi_3}e^{-i\alpha L_4}e^{-in\phi_4}e^{i\alpha L_5}e^{in\phi_6}\right)^{m},
\end{align}
where $m\in\z$ is quantized because of the compactification condition on $\phi$. When the system is put on a torus (see Fig.~\ref{fig: non-contractible loops}), there are two more symmetries along the non-contractible loops:
\beq
\begin{split}
W_\eta(m)&=\Big(\prod_{i\in A}e^{-i\alpha L_i}\prod_{i\in B}e^{i\alpha L_i}\Big)^m,\\
W_\gamma(m)&=\Big(\prod_{i\in A}e^{in\phi_i}\prod_{i\in B}e^{-in\phi_i}\Big)^m.
\end{split}
\eeq
These non-contractible symmetry operators satisfy $W_\eta(1) W_\gamma(1) W^{-1}_\eta(1) W^{-1}_\gamma(1)= e^{2i n \alpha}$. When $e^{2in\alpha}$ is an $N$-th root of unity, each energy level of any Hamiltonian with such a symmetry must be $N$-fold degenerate on a torus. By contrast, when $e^{i\alpha}$ is not a root of unity, each energy level of any Hamiltonian with such a symmetry must be infinitely degenerate on a torus.

A Kitaev-type Hamiltonian with this prescribed symmetry can be taken as
\beq
H=\sum_{\mu=X,Y,Z}\sum_{\langle i,j\rangle=\mu}f_{\mu},
\eeq
where the summation runs over all links connecting nearest neighbors, the labels of the bonds are defined in Fig.~\ref{fig:label_of_links}, and
\beq
\begin{split}
    f_{X}&=f_{X}(\phi_{i}+\phi_{j}),\\
    f_{Y}&=f_{Y}(\alpha L_{i}+n\phi_{i}+\alpha L_{j}+n\phi_{j}), \\
    f_{Z}&=f_{Z}(L_{i}+L_{j}),
\end{split}
\eeq
with $f_{X, Y, Z}$ functions that are bounded from below, and, moreover, $f_X$ a $2\pi$-periodic function and $f_{Y}$ a $2\pi n$-periodic function. This model cannot be solved analytically due to its interacting nature. However, note that the operator $W_l$
has a continuous spectrum obtained by shifting the $\phi$'s, hence we also expect any Hamiltonian with such a $\z$ 1-form symmetry has a spectrum of the form of Fig.~\ref{fig: possibly spurious spectrum}. This $\z$ 1-form symmetry is thus expected to not flow to a topological 1-form symmetry at low energies.

\bibliography{ref.bib}

\end{document}